\documentclass[preprints,review,accept,moreauthors,pdftex]{mdpi}
\pdfoutput=1

\firstpage{1} 
\pubvolume{1}
\issuenum{1}
\articlenumber{0}
\pubyear{2022}
\copyrightyear{2022}
\externaleditor{Academic Editor:}
\datereceived{} 
\dateaccepted{} 
\datepublished{} 
\hreflink{https://doi.org/} % If needed use \linebreak
\Title{Overview of the Observing System and Initial Scientific Accomplishments of the East Asian VLBI Network (EAVN)}

\TitleCitation{Overview of the Observing System and Initial Scientific Accomplishments of the East Asian VLBI Network (EAVN)}

\Author{Kazunori Akiyama $^{1,2,3}$, %% MIT, NAOJ Miz, BHI
Juan-Carlos Algaba $^{4}$, %% U. Malaya
Tao An $^{5,6}$, %% SHAO, KLRA
Keiichi Asada $^{7}$, %% ASIAA
Kitiyanee Asanok $^{8}$, %% NARIT
Do-Young Byun $^{9,10}$, %% KASI, UST
Thanapol Chanapote $^{8}$, %% NARIT
Wen Chen $^{11}$, %% YNAO
Zhong Chen $^{5}$, %% SHAO
Xiaopeng Cheng $^{9}$
% ...
on behalf of the East Asian VLBI Network Collaboration
}

\AuthorNames{
East Asian VLBI Network Collaboration, Kiyoaki Wajima
}

\AuthorCitation{Akiyama, K.; Algaba, J.-C.; An, T.; Asada, K.; Asanok, K.; Byun, D.-Y.; Chanapote, T.; Chen, W.; Chen, Z.; Cheng, X.; et~al.}

\address{%
$^{1}$ \quad Massachusetts Institute of Technology Haystack Observatory, Westford, MA 01886, USA \\ %% MIT
$^{2}$ \quad Mizusawa VLBI Observatory, National Astronomical Observatory of Japan, Iwate 023-0861, Japan \\ %% NAOJ Miz
$^{3}$ \quad Black Hole Initiative, Harvard University, Cambridge, MA 02138, USA \\ %% BHI
$^{4}$ \quad Department of Physics, Faculty of Science, Universiti Malaya, Kuala Lumpur 50603, Malaysia \\ %% U. Malaya
$^{5}$ \quad Shanghai Astronomical Observatory, Chinese Academy of Sciences, Shanghai 200030, China \\ %% SHAO
$^{6}$ \quad Key Laboratory of Radio Astronomy, Chinese Academy of Sciences, Nanjing 210008, China \\ %% KLRA
$^{7}$ \quad Institute of Astronomy and Astrophysics, Academia Sinica, Hilo, HI 96720, USA \\ %% ASIAA
$^{8}$ \quad National Astronomical Research Institute of Thailand (Public Organization), Chiangmai 50180, Thailand \\ %% NARIT
$^{9}$ \quad Korea Astronomy and Space Science Institute, Yuseong-gu, Daejeon 34055, Republic of Korea \\ %% KASI
$^{10}$ \hspace{-1mm}\quad Department of Astronomy and Space Science, University of Science and Technology, Yuseong-gu, Daejeon 34113, Republic of Korea \\ %% UST
$^{11}$ \hspace{-1mm}\quad Yunnan Observatories, Chinese Academy of Sciences, Kunming 650011, China \\ %% YNAO
% ...
}

\corres{Correspondence: wajima@kasi.re.kr}

\abstract{
The East Asian VLBI Network (EAVN) is an international VLBI facility in East Asia and is operated under mutual collaboration between East Asian countries, as well as part of Southeast Asian and European countries.
EAVN currently consists of 16 radio telescopes and three correlators located in China, Japan, and Korea, and is operated mainly at three frequency bands, 6.7, 22, and 43~GHz with the longest baseline length of 5078~km, resulting in the highest angular resolution of 0.28 milliarcseconds at 43~GHz.
One of distinct capabilities of EAVN is multi-frequency simultaneous data reception at nine telescopes, which enable us to employ the frequency phase transfer technique to obtain better sensitivity at higher observing frequencies.
EAVN started its open-use program in the second half of 2018,  providing a total observing time of more than 1100 hours in a year.
EAVN fills geographical gap in global VLBI array, resulting in enabling us to conduct contiguous high-resolution VLBI observations.
EAVN has produced various scientific accomplishments especially in observations toward active galactic nuclei, evolved stars, and star-forming regions.
These activities motivate us to initiate launch of the 'Global VLBI Alliance’ to provide an opportunity of VLBI observation with the longest baselines on the earth.
}


\begin{document}

%%%%%%%%%%%%%%%%%
%%% Section 1 %%%
%%%%%%%%%%%%%%%%%

\section{Introduction}
\label{sec:Introduction}

The very long baseline interferometry (VLBI) technique enables us to obtain extremely high angular resolution of milliarcsecond (mas) to microarcsecond ($\mu$as) scales \cite{Thompson17}.
In VLBI, higher angular resolution can be achieved by employing longer baseline length between each telescope or by conducting an observation at higher observing frequency.
Longer baseline lengths can be obtained with global-scale VLBI array such as, for example, the European VLBI Network (EVN) \cite{Zensus15}, the Long Baseline Array (LBA) in Australia \cite{Edwards15}, the Very Long Baseline Array (VLBA) \cite{Napier94}, and the Global Millimeter-VLBI Array \mbox{(GMVA) \cite{Hodgson12}}, or can also be achieved with so-called `space VLBI' technique by employing one (or more) satellite(s) and conducting a VLBI observation using a satellite and ground-based radio telescopes.
Space VLBI observations were realized by the VLBI experiment with the Tracking and Data Relay Satellite System (TDRSS) \cite{Levy86,Linfield89,Linfield90}, the VLBI Space Observatory Programme (VSOP) with the HALCA spacecraft \cite{Hirabayashi98,Hirabayashi00}, and the RadioAstron project \cite{Kardashev13}.
On the other hand, one of the ultimate goals in the angular resolution with higher observing frequency was accomplished by a series of VLBI observations of a black hole shadow in our Galactic center Sgr A$^*$ and a nearby active galaxy M87 at 230~GHz led by the Event Horizon Telescope (EHT) project with its angular resolution of about 25~$\mu$as \cite{EHT19a,EHT22a}.

VLBI is an important tool for studying Galactic and extragalactic astrophysical phenomena mainly on the basis of mas- to $\mu$as-scale images \cite{Verschuur88}, while it also has various applications in geophysics, such as, for example, measurement of the Earth orientation parameter (EOP; \cite{Schuh12,Nothnagel17} and references therein), determination of the International Celestial Reference Frame (ICRF; \cite{Ma98,Fey15,Charlot20,Witt22}) and the International Terrestrial Reference Frame (\mbox{ITRF; \cite{Altamimi16}} and references therein), and study on the Earth's tectonic plate (\cite{Sovers98} and references therein).
Capabilities of these applications can be developed in terms of an accuracy of measurement by conducting high-cadence monitoring with high angular resolution by employing intercontinental VLBI observations (e.g., \cite{Ghaderpour21}).
These examples thus indicate the importance of close international collaboration in VLBI.

Since the first success of a VLBI experiment in 1967 \cite{Broten67,Kellermann01}, a lot of domestic and international VLBI networks have been developed and it is also the case in East Asian countries.
In Japan, pioneering domestic VLBI experiments were carried out using Kashima 34~m and Nobeyama 45~m telescopes called as `the Kashima--Nobeyama Interferometer' (KNIFE) at 22 and 43~GHz early in 1990s \cite{Miyoshi93}.
Nobeyama also joined the intercontinental millimeter-wavelength VLBI experiment at 100~GHz \cite{Baath92}.
Series of VLBI experiments with KNIFE led launch of the first Japanese-domestic VLBI array entitled `the Japanese VLBI Network' (J-Net) in 1994 consisting of four telescopes (Kashima 34~m, Nobeyama 45~m, Mizusawa 10~m, and Kagoshima 6~m).
J-Net made couple of new findings in its kinematics with observations of water-vapor masers in and around star-forming regions and protostars (e.g., \cite{Omodaka99,Imai99}) and evolved stars (e.g., \cite{Imai97,Ishitsuka01}).
J-Net also took an important role in a Galactic-plane survey of extragalactic radio sources for a positional reference of Galactic maser sources \cite{Honma00a} as a pathfinder of the VLBI Exploration of Radio Astrometry (VERA) \cite{Honma00b,Kobayashi03}.
VERA is also a key partner of `the (renewed) Japanese VLBI Network' (JVN) \cite{Doi06} consisting of several telescopes operated by universities and institutes in which each telescope is located (Tomakomai 11~m operated by Hokkaido University \cite{Sorai08}; Hitachi 32~m and Takahagi 32~m operated by Ibaraki University \cite{Yonekura16}; Tsukuba 32~m operated by Geospatial Information Authority of Japan (GSI) \cite{Kokado08}; Kashima 34~m operated by the National Institute of Information and Communications Technology (NICT), Usuda 64~m operated by Japan Aerospace Exploration Agency (JAXA), Gifu 11~m operated by Gifu University, and Yamaguchi 32~m operated by Yamaguchi University \cite{Fujisawa02})\endnote{Tomakomai 11~m, Tsukuba 32~m, and Kashima 34~m telescopes have already been deconstructed.}.

China participated in international VLBI experiments in 1980s.
The first radio telescope in China was constructed at Sheshan, a suburban area of Shanghai, with its diameter of 25~m in 1987 \cite{Liang89}.
Sheshan is a telescope which participated in EVN as a full member at an early stage of EVN in 1990.
After that many telescopes in China started participation in EVN, Nanshan 25~m in 1994 (main reflector of the telescope has been refurbished with its diameter of 26~m in 2015) \cite{Zhang19}, Kunming 40~m in 2009 \cite{Hao10}, and Tianma 65~m in \mbox{2015 \cite{Bolli19}}.
These telescopes constitute `the Chinese VLBI Network' (CVN) \cite{Li08} with the longest baseline length of 3300~km between Sheshan and Nanshan telescopes.
In addition to astronomical VLBI observations, CVN is utilized for determining an orbit of a series of Chang'E lunar explorers (\cite{Liu21} and references therein) and the Tianwen-1 Mars \mbox{explorer \cite{Yang22}}, and for making a precise position determination of landers for these spacecrafts \cite{Klopotek19}.

The first radio telescope in Korea, Taeduk Radio Astronomy Observatory (TRAO) 14-m telescope, was constructed in 1986 with its major observing frequencies of around and greater than 100~GHz \cite{Jeong19}.
TRAO 14~m conducted the first astronomical VLBI observation with Nobeyama 45~m at 86~GHz in 2001 and successful fringe detection of SiO $v$ = 1, $J$ = 2--1 maser emission was made toward an evolved star VY Canis Majoris \cite{Shibata04}.
This experiment was an important demonstration for the construction of a domestic VLBI array `the Korean VLBI Network' (KVN) from \mbox{2001 \cite{Wajima05,Lee11,Lee14}}.
One of the most unique capabilities of KVN is simultaneous multifrequency data receiving system for four observing frequencies (22, 43, 86, and 129~GHz), which is the world's first system equipped in a radio \mbox{telescope \cite{Han13}}.
Antenna's basic performance of KVN telescopes, including the pointing accuracy and beam pattern measurement, was confirmed using a temporary receiver at 100~GHz before starting the open-use program of KVN in 2013 \cite{Kim11}.
KVN started participation in EVN as an associate member in 2014.

As mentioned above, domestic VLBI arrays in each country in East Asia have their own geographical uniqueness.
Japan (VERA and JVN) shows a wide distribution of a lot of radio telescopes in the country with the maximum baseline length of 2300~km, whereas KVN is a compact array with the range of its baseline length of 305--480~km, resulting in making very high-sensitive array with shorter baseline lengths.
Some of CVN telescopes locate at the near-Easternmost and near-Westernmost parts of its very big country.
The features found in each VLBI array are complementary in terms of an angular resolution and an imaging quality.
Those situations motivate us to launch a new international VLBI array in East Asia called `the East Asian VLBI Array' (EAVN) \cite{Wajima16,An18}.
The very early concept of EAVN, however, was discussed in the mm-VLBI workshop held in 2003 in China and the committee of the East Asian VLBI Consortium was established at the 6th East Asian Meeting on Astronomy (EAMA) in 2004 \cite{Inoue05}.
This consortium was transferred to a task force for EAVN as one of working groups under the East Asian Core Observatories Association (EACOA)\endnote{\url{https://www.eacoa.net/}, accessed on 1 December 2022} in 2005 and started its practical activity from 2013 including creation of a governing body of EAVN and evaluation of EAVN's array capability.
Prior to launch of EAVN, bilateral VLBI array between Korea and Japan, the KVN and VERA Array (KaVA; e.g., \cite{Matsumoto14,Niinuma14}), has been started its operation in 2013 and its open-use program in 2014.
KaVA has conducted more than 650 observing sessions with the total observing time of about 4700 hours by the middle of 2018.
After completing a series of fringe test observations \cite{Wajima16}, evaluation of data calibration capability \cite{Cho17}, and pioneering scientific experiments with EAVN (e.g., \cite{Fujisawa14,Sugiyama16}), open-use program of EAVN has been launched in the second half of 2018.

This paper provides basic information on the overall observing system of EAVN and reviews major scientific results obtained by EAVN to date.
EAVN is a unique VLBI array not only for filling a geographical gap between existing VLBI arrays on the Earth but also for providing an opportunity to conduct an observation using the multi-frequency simultaneous data reception system with global-scale VLBI for the first time, both of which are described in Section~\ref{sec:Overview of the Array}.
Astronomers all over the world can utilize these capabilities through the EAVN open-use program, which is described in Section~\ref{sec:Summary of the EAVN Open-Use Program}.
Thanks to the unique capabilities mentioned above, EAVN has made scientific findings toward active galactic nuclei (AGNs), evolved stars, and star-forming regions, as well as various Galactic and extragalactic radio sources, on the basis of high-cadence VLBI monitoring and simultaneous data reception of two or more radio signals at cm- to mm-wavelength regime, which are shown in Section~\ref{sec:Scientific Accomplishments by EAVN}.
EAVN is also grown as one of the key VLBI arrays in terms of future international collaboration and those activities are summarized in Section~\ref{sec:Future Growth into the Global VLBI}.
Note that part of scientific results shown in this paper are derived with not only EAVN but also bilateral VLBI collaboration between East Asian countries, such as KaVA.

%%%%%%%%%%%%%%%%%
%%% Section 2 %%%
%%%%%%%%%%%%%%%%%

\section{Overview of the Array}
\label{sec:Overview of the Array}

This section describes an overview of the observation system of EAVN.
The system is always updated and its detailed specifications can be found in `the EAVN Status Report' on the EAVN website\endnote{\url{https://eavn.kasi.re.kr/}, accessed on 1 December 2022}.

%%%%%%%%%%%%%%%%%%%
%%% Section 2-1 %%%
%%%%%%%%%%%%%%%%%%%

\subsection{Telescope}
\label{subsec:Telescope}

\begin{figure}[H]
\includegraphics[width=12.5 cm]{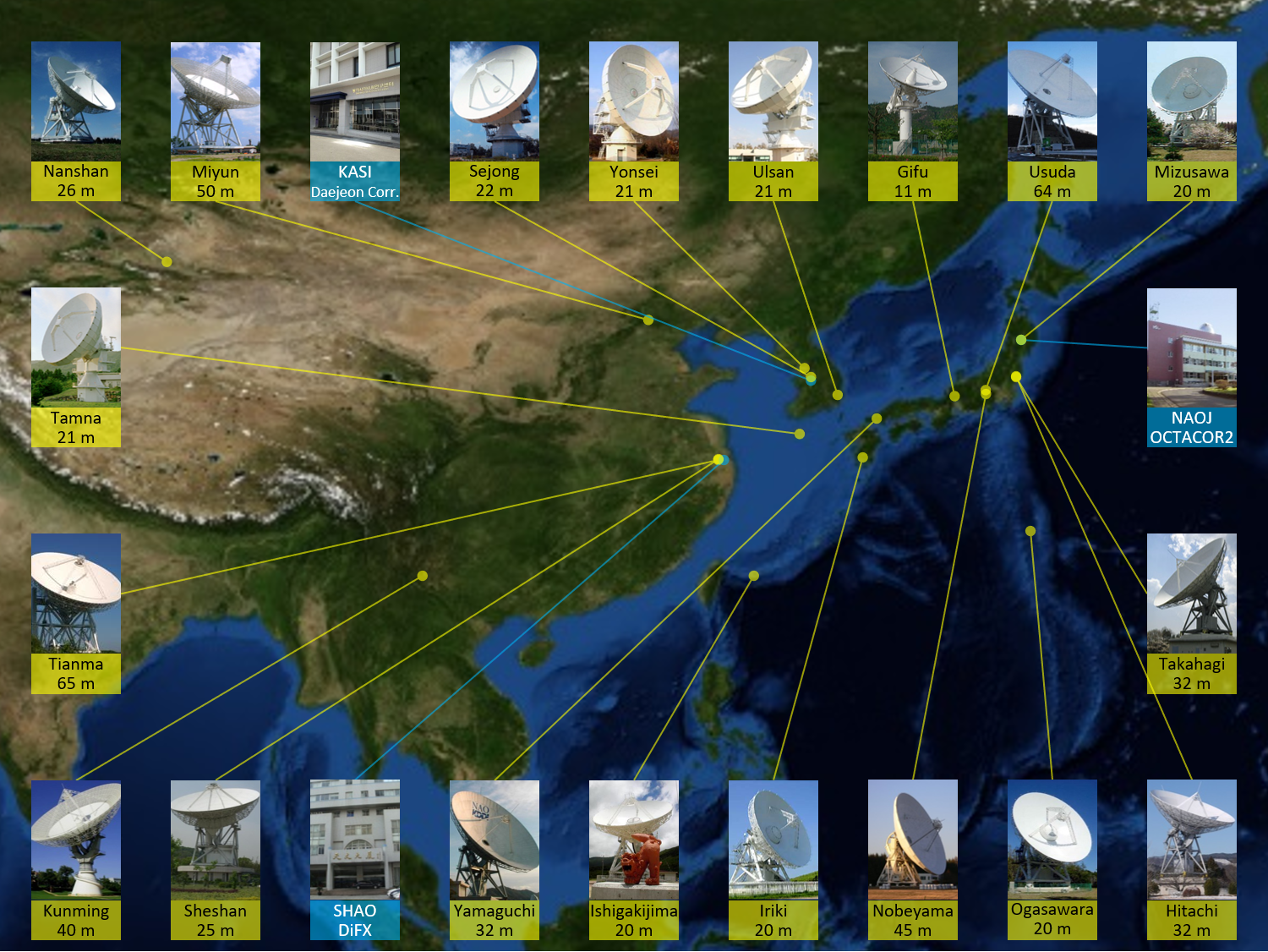}
\caption{Concept map of EAVN including 16 active telescopes listed in Table~\ref{tbl:EAVN-List}, three potential
telescopes (Miyun 50~m in China, Gifu 11~m, and Usuda 64~m in Japan), and three correlator sites in China, Japan,
and Korea.
Photographs of each radio telescope and correlator site are overlaid on `the Blue Marble' image (credit of the ground image: NASA Earth Observatory (\url{https://earthobservatory.nasa.gov/}, accessed on 1 December 2022).
\label{fig:EAVN-map}}
\end{figure}

Figure~\ref{fig:EAVN-map} shows the location map for each participating telescope in EAVN which consists of 16 radio telescopes and three correlator sites located in each East Asian country (Daejeon in Korea, Shanghai in China, and Mizusawa in Japan).
EAVN composes of several independent VLBI arrays and radio telescopes.
VERA consists of four telescopes (Iriki, Ishigakijima, Mizusawa, and Ogasawara) and is a core VLBI array in Japan.
Other Japanese telescopes except Nobeyama are joint partners for JVN.
Four Chinese telescopes (Kunming, Nanshan, Sheshan, and Tianma) constitute CVN.
KVN consists of three telescopes (Seoul, Ulsan, and Jeju).
Sejong 22-m telescope has been constructed in 2013 with space geodetic survey being as a main task, while the telescope joined the open-use program for both KVN and EAVN in 2022.

%%%%%%%%%%%%%%%%
%%% Figure 1 %%%
%%%%%%%%%%%%%%%%

\begin{figure}[H]
\includegraphics[width=13.5 cm]{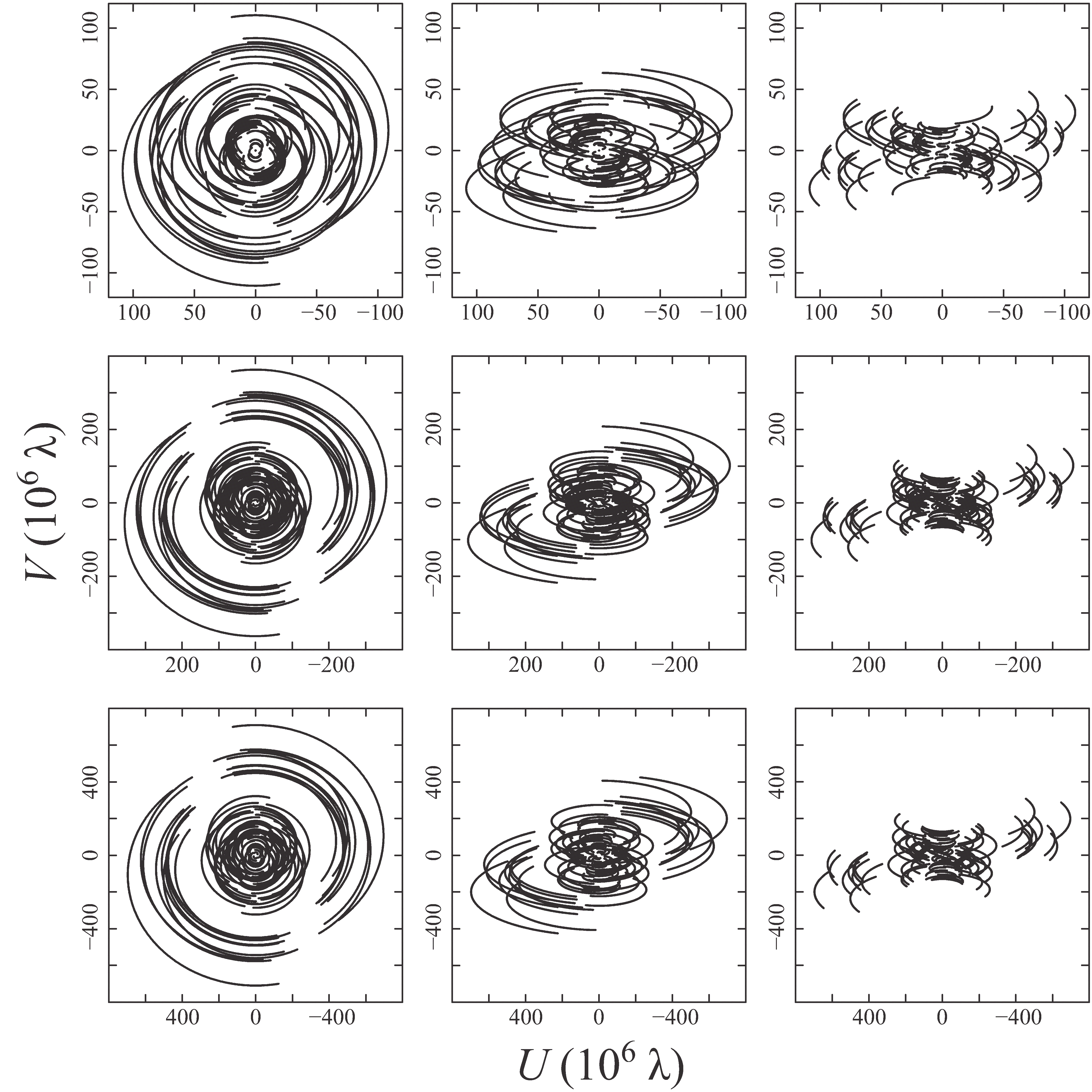}
\caption{Examples of the ($u$, $v$) coverage for an EAVN observation with full array configuration at 6.7 (upper panels), 22 (middle panels), and 43~GHz (lower panels) with the source's declination of $+60^{\circ}$ (left panels), $+20^{\circ}$ (center panels), and $-20^{\circ}$ (right panels).
$\lambda$ represents the observing wavelength.
Total observing duration of 10 hours and the antenna's lower elevation limit of $15^{\circ}$ are assumed for all cases.
\label{fig:EAVN-UV}}
\end{figure}

Figure~\ref{fig:EAVN-UV} shows examples of the spatial frequency coverage (($u$, $v$) coverage) of EAVN at each observing frequency.
The maximum baseline length of EAVN is composed of Nanshan and VERA-Ogasawara stations with its baseline length of 5078~km, corresponding to an angular resolution of 1.82, 0.55, and 0.28~mas at 6.7, 22, and 43~GHz, respectively.

%%%%%%%%%%%%%%%%
%%% Figure 2 %%%
%%%%%%%%%%%%%%%%

Table~\ref{tbl:EAVN-List} indicates basic observing parameters for participating telescopes of EAVN.
EAVN is operated using 11, 12, and 11 telescopes at three frequency bands, 6.7, 22, and 43~GHz, respectively.
The number of participating telescopes shown above includes Nanshan at 6.7 and 43~GHz in which performance of the cooled receiver at these frequencies is under evaluation.
The system equivalent flux density (SEFD) shown in Table~\ref{tbl:EAVN-List} is defined as follows;
\begin{equation}
SEFD = \frac{2 k_{\rm B} T_{{\rm sys}, \nu}}{A_{{\rm e}, \nu}},
\label{eqn:SEFD}
\end{equation}
where $k_{\rm B}$ is Boltzmann's constant, $T_{{\rm sys}, \nu }$ and $A_{{\rm e}, \nu}$ are the system noise temperature and the effective aperture area at an observing frequency $\nu$, respectively.

%%%%%%%%%%%%%%%
%%% Table 1 %%%
%%%%%%%%%%%%%%%

%\begin{adjustwidth}{-\extralength}{0cm}
%\centering 
%% If there is a figure in wide page, please release command \centering
\begin{table}[H]
\caption{Participating arrays and telescopes in EAVN.
\label{tbl:EAVN-List}}
\tablesize{\endnotesize}
\setlength{\tabcolsep}{1.4mm}{\begin{tabular}{llcccccc}
\toprule
\textbf{Country} & \textbf{Array/Telescope}	& \textbf{Enrollment} \boldmath{$^a$} & \textbf{Diameter [m]} & \multicolumn{3}{c}{\textbf{SEFD \boldmath{$^b$} [Jy]}} \\
                 &                          &                     &                       & \textbf{6.7~GHz} & \textbf{22~GHz} & \textbf{43~GHz} \\
\midrule
China            & Kunming                  & 2021                & ~~~~~~~40             &  ~~307      &--      &--        \\
                 & Nanshan                  & 2018                & ~~~~~~~26             & *** $^d$    & ~~364  & *** $^d$ \\
                 & Sheshan                  & 2021                & ~~~~~~~25             &   1740      &--      &--        \\
                 & Tianma                   & 2018                & ~~~~~~~65             & ~~~~55      & ~~100  & ~~122    \\
Japan            & Hitachi                  & 2021                & ~~~~~~~32             &  ~~158      &--      &--        \\
                 & Nobeyama                 & 2018                & ~~~~~~~45             &--           & ~~285  & ~~655    \\
                 & Takahagi                 & 2020                & ~~~~~~~32             &--           & ~~343  &--        \\
                 & VERA                     & 2018                & $4 \times 20$         &   2155      &  2110  &  4393    \\
                 & Yamaguchi                & 2021                & ~~~~~~~32             &  ~~286      &--      &--        \\
Korea            & KVN                      & 2018                & $3 \times 21$         &   4241 $^c$ &  1328  &  1992    \\
                 & Sejong                   & 2022                & ~~~~~~~22             &--           &  1231  &  2055    \\
\bottomrule
\end{tabular}}
%\end{adjustwidth}

\noindent{\footnotesize{Note. $^a$ Year of enrolment in EAVN. $^b$ The system equivalent flux density (SEFD) for VERA and KVN is a value for a single antenna. $^c$ Only Ulsan station can conduct an observation at 6.7~GHz. $^d$ The system performance is under evaluation.}}
\end{table}

%%%%%%%%%%%%%%%%%%%
%%% Section 2-2 %%%
%%%%%%%%%%%%%%%%%%%

\subsection{Overview of Data Acquisition Flow}
\label{subsec:Overview of Data Acquisition Flow}

\begin{figure}[H]	
\includegraphics[width=13.5 cm]{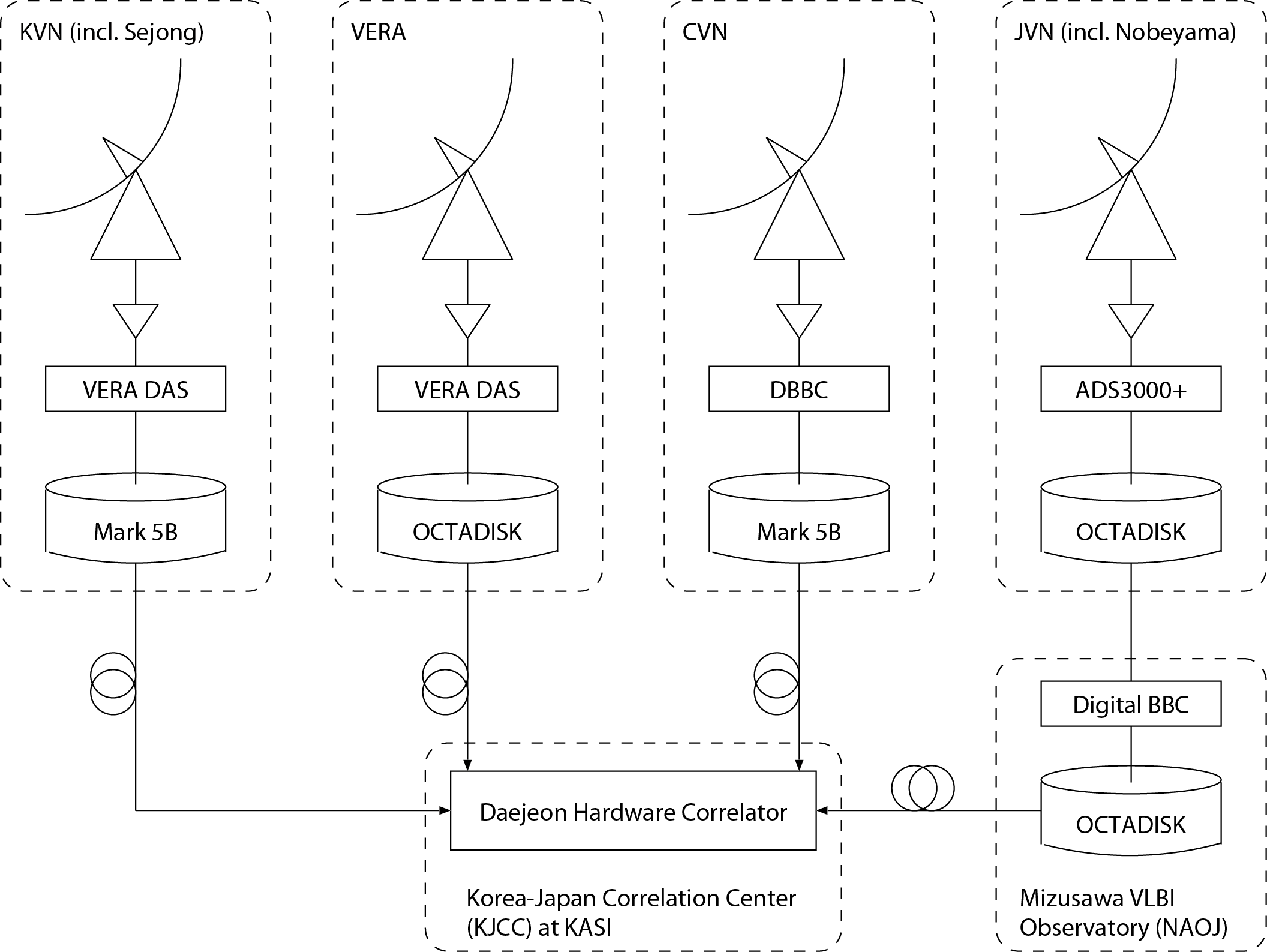}
\caption{Schematic diagram of data flow of current EAVN observations. `DAS' represents data acquisition system.
\label{fig:EAVN-Data-Flow}}
\end{figure}

Figure~\ref{fig:EAVN-Data-Flow} shows a schematic diagram of data flow of an EAVN observation.
Radio signal received is digitized and formatted using digital signal processing (DSP) system, and recorded with a disk storage at each telescope.
Recorded data are transferred to the Korea--Japan Correlation Center (KJCC), in which the main correlator for EAVN is equipped, via FTP.
Data at Nobeyama and three JVN stations (Hitachi, Takahagi, and Yamaguchi) are recorded by the OCTADISK recorder \cite{Oyama12} and those data are once transferred to the digital baseband converter (DBBC) at Mizusawa VLBI Observatory to perform digital filtering before sending them to KJCC.
Details of this process are mentioned in Section~\ref{subsec:Digital Signal Processing and Recorder}, while note that the process becomes unnecessary for future EAVN observations with the data recording rate of 2~Gbps or higher.

%%%%%%%%%%%%%%%%
%%% Figure 3 %%%
%%%%%%%%%%%%%%%%

%%%%%%%%%%%%%%%%%%%
%%% Section 2-3 %%%
%%%%%%%%%%%%%%%%%%%

\subsection{Frontend}
\label{subsec:Frontend}

Receiver noise temperature of EAVN telescopes for each observing frequency is shown in Table~\ref{tbl:EAVN-Backend}.
Cooled HEMT receiver for all observing frequencies of EAVN is equipped at all EAVN telescopes, except a 6.7~GHz receiver at all VERA stations, KVN-Ulsan, and Sheshan being operated under a room temperature.
Data reception with dual polarization is available for all receivers except a 6.7~GHz receiver at VERA and KVN-Ulsan and a 43~GHz receiver at Nobeyama being available for only left-hand circular polarization.

One of the most unique capabilities of EAVN is a multi-frequency data receiving system in which an observation data can be obtained at two or more frequencies simultaneously.
Currently such system is installed at nine stations, KVN (22, 43, 86, and 129~GHz), VERA (22 and 43~GHz), Sejong (22 and 43~GHz), and Nobeyama (22, 43, and 86~GHz).
This unique capability enables us to obtain better baseline sensitivity by employing the `frequency phase transfer' (FPT) technique \cite{Algaba15,Zhao18}, and to determine a relative position between target and reference source by employing the `source-frequency phase-referencing' (SFPR) \mbox{technique \cite{Dodson14,Zhao19}}.
Details of the system installed at Nobeyama, the HINOTORI (Hybrid Installation Project in Nobeyama, Triple-band Oriented) system, is provided by another paper \cite{Okada20}.

It is also important not only to equip a high-performance receiver but to provide a source catalog optimized for an observation at 22~GHz or higher, which are main observing frequency with EAVN, in order to raise the detectability of target sources and to provide an appropriate environment for calibration of the data obtained by EAVN since VLBI calibrators at higher frequency of greater than 20~GHz are less known compared to those at lower frequency. 
A series of VLBI calibrator survey containing more than 1500 extragalactic radio sources with its declination of higher than $-32.5^{\circ}$ are carried out using KVN at 22 and 43~GHz, resulting in providing an important information for calibration of the flux density with KaVA and EAVN \cite{Lee17}.

%%%%%%%%%%%%%%%
%%% Table 2 %%%
%%%%%%%%%%%%%%%

%\centering %% If there is a figure in wide page, please release command \centering
\begin{table}[H]
\caption{Performance of the receiver and the data acquisition system at each EAVN station.
\label{tbl:EAVN-Backend}}
\tablesize{\endnotesize}

\begin{adjustwidth}{-\extralength}{0cm}
%\centering %% If there is a figure in wide page, please release command \centering
\setlength{\tabcolsep}{4.3mm}{\begin{tabular}{llcccccc}
\toprule
\textbf{Array/Telescope} & \multicolumn{3}{c}{\boldmath{$T_{\rm RX}$}} & \textbf{DSP} & \textbf{Recorder} \\
                         & \textbf{6.7~GHz} & \textbf{22~GHz} & \textbf{43~GHz}                 &              & \\
\midrule
Kunming                  & ~~20     & ---    & ---                    &  DBBC, CDAS  & Mark 5B, Mark 6   \\
Nanshan                  & *** $^a$ & 15     & *** $^a$               &  DBBC, CDAS  & Mark 5B, Mark 6   \\
Sheshan                  & 100      & ---    & ---                    &  DBBC, CDAS  & Mark 5B, Mark 6   \\
Tianma                   & ~~20     & 35     & ~~50                   &  DBBC, CDAS  & Mark 5B, Mark 6   \\
Hitachi                  & ~~20     & ---    & ---                    &  ADS3000+    & OCTADISK          \\
Nobeyama                 & ---      & 85     & 110                    &  ADS3000+    & OCTADISK          \\
Takahagi                 & ---      & 30     & ---                    &  ADS3000+    & OCTADISK          \\
VERA                     & 100      & 50     & ~~90                   &  VERA, OCTAD & OCTADISK          \\
Yamaguchi                & ~~20     & ---    & ---                    &  ADS3000+    & OCTADISK          \\
KVN                      & 300      & 40     & ~~50                   &  VERA, OCTAD & Mark 5B, Mark 6, OCTADISK \\
Sejong                   & ---      & 30     & ~~90                   &  DBBC        & Mark 6            \\
\bottomrule
\end{tabular}}
\end{adjustwidth}
\noindent{\footnotesize{Note. $^a$ The system performance is under evaluation.}}
\end{table}

%%%%%%%%%%%%%%%%%%%
%%% Section 2-4 %%%
%%%%%%%%%%%%%%%%%%%

\subsection{Digital Signal Processing and Recorder}
\label{subsec:Digital Signal Processing and Recorder}

Table~\ref{tbl:EAVN-Backend} shows an equipment for DSP and a recorder used at each telescope.
EAVN is currently operated with the data recording rate of 1024~Mbps, corresponding to the total bandwidth of 256~MHz for 2-bit sampling.
Due to a difference in the historical background of each VLBI array constituting EAVN, DSP and recorder installed are somewhat different.

To keep compatibility of digital data sequence between KVN and VERA for conducting joint observations with KaVA, similar type of digital data acquisition system \cite{Iguchi05} is introduced at KVN stations \cite{Oh11}.
CVN telescopes, on the other hand, introduce the DBBC system partly because they have started participation in EVN prior to EAVN.
Another type of DSP system, the Chinese VLBI Data Acquisition System (CDAS), is developed uniquely in China and installed at all CVN telescopes \cite{Zhang10}.

Observation data at four Japanese telescopes (Hitachi, Nobeyama, Takahagi, and Yamaguchi) are digitized by the ADS3000+ sampler \cite{Takefuji10} and recorded using the OCTADISK system \cite{Oyama16}, both of which were developed uniquely in Japan, while ADS3000+ does not have a same digital filter mode to that of VERA DAS and DBBC.
Data obtained at these telescopes are thus recorded by OCTAD with the data output rate of 2~Gbps and transferred to the Mizusawa VLBI Observatory to convert the data output rate (1~Gbps) and the channel assignment of the data suitable for correlation with other telescopes' data.
Note that EAVN is planning to move to regular operation with the data recording rate of 2~Gbps or higher in the near future, data processing at Mizusawa, shown in Figure~\ref{fig:EAVN-Data-Flow}, will thus become unnecessary.

%%%%%%%%%%%%%%%%%%%
%%% Section 2-5 %%%
%%%%%%%%%%%%%%%%%%%

\subsection{Correlator}
\label{subsec:Correlator}

EAVN is operated using correlators located in China, Japan, and Korea, as shown in Table~\ref{tbl:EAVN-Correlator}.
All correlators are utilized for examining data quality and various observing modes of EAVN, whereas all observation data on the open-use program are correlated using the Daejeon Correlator, which is hardware-based, at the headquarters of Korea Astronomy and Space Science Institute (KASI).
DiFX correlator at Daejeon and Shanghai is mainly used for correlation of data obtained by KVN and CVN, respectively.
The 'SoftCos' correlator was assembled by National Astronomical Observatory of Japan (NAOJ) \cite{Oyama16} on the basis of the GICO3 software correlator as a main engine which was developed by \mbox{NICT \cite{Kimura02}.}

%%%%%%%%%%%%%%%
%%% Table 3 %%%
%%%%%%%%%%%%%%%

\begin{table}[H] 
\caption{Correlators used for EAVN observations.
\label{tbl:EAVN-Correlator}}
\setlength{\tabcolsep}{9.05mm}{\begin{tabular}{llcccccc}
\toprule
\textbf{Site}      & \textbf{Correlator}      & \textbf{Reference}      \\
\midrule
Daejeon (KASI)     & Daejeon Correlator, DiFX & \cite{Lee15,Deller07}   \\
Shanghai (SHAO)    & DiFX                     & \cite{Deller07,Jiang18}         \\
Mizusawa (NAOJ)    & SoftCos                  & \cite{Kimura02,Oyama16} \\
\bottomrule
\end{tabular}}
\end{table}

Table~\ref{tbl:Daejeon Correlator} represents basic specifications of the Daejeon Correlator.
Results of performance evaluation of the Daejeon Correlator can be found in the literature \cite{Lee15}.
The Daejeon Correlator has 16 data input ports, named as the antenna unit, corresponding to a single polarization (right-hand circular polarization (RHCP) or left-hand circular polarization (LHCP)) at a single frequency for one telescope.
This results in the maximum number of telescopes for data correlation at a single path being 16 for an observation with a single polarization at a single frequency, or 8 for an observation with dual polarization at a \mbox{single frequency.}

%%%%%%%%%%%%%%%
%%% Table 4 %%%
%%%%%%%%%%%%%%%

\begin{table}[H] 
\caption{Basic specifications of the Daejeon Correlator.
\label{tbl:Daejeon Correlator}}
%%% \tablesize{} %% You can specify the fontsize here, e.g., \tablesize{\endnotesize}. If commented out \small will be used.
\setlength{\tabcolsep}{8.4mm}{\begin{tabular}{llcccccc}
\toprule
Maximum number of stations & 16                         \\
Maximum delay compensation & $\pm 36000$~km            \\
Maximum fringe tracking    & 1.075~kHz                  \\
Maximum data input rate    & 2048~Mbps                  \\
Maximum number of spectral points per IF channel & 8192 \\
Integration time           & 25.6~ms--10.24~s         \\
\bottomrule
\end{tabular}}
\end{table}

%%%%%%%%%%%%%%%%%
%%% Section 3 %%%
%%%%%%%%%%%%%%%%%

\section{Summary of the EAVN Open-Use Program}
\label{sec:Summary of the EAVN Open-Use Program}

As mentioned in Sections~\ref{sec:Introduction} and \ref{sec:Overview of the Array}, a domestic VLBI array is operated in each East Asian country, JVN and VERA in Japan, CVN in China, and KVN in Korea, and the open-use program is also carried out by each VLBI array.
Although each VLBI array has its own advantage in both scientific and technical point of view, it is a natural expansion to constitute a larger VLBI network by combining adjacent domestic VLBI arrays.
On the basis of scientific and technical accomplishments with KaVA and initial performance evaluation with EAVN shown in Section~\ref{sec:Introduction}, the EAVN open-use program has been started from the second half of 2018.

EAVN consists of 16 telescopes in East Asian countries, while the observation time for EAVN to be provided by each telescope time is different between each other, as shown in Table~\ref{tbl:EAVN Telescope Time}.
Proposers can choose any combination of an array configuration so that the configuration becomes a suitable one for their scientific purposes, although KaVA shall participate in all EAVN observations in principle.

%%%%%%%%%%%%%%%
%%% Table 5 %%%
%%%%%%%%%%%%%%%

\begin{table}[H] 
\caption{Total observation time of EAVN provided by each telescope for the open-use program in the 2022B semester.
\label{tbl:EAVN Telescope Time}}
\setlength{\tabcolsep}{8.5mm}{\begin{tabular}{llcccccc}
\toprule
\textbf{Country} & \textbf{Array/Telescope} & \textbf{Observing Time for EAVN [h]}  \\
\midrule
China            & Kunming                  & ~~50 \\
                 & Nanshan                  &  150 \\
                 & Tianma/Sheshan           &  150 \\
Japan            & VERA                     &  500 \\
                 & Hitachi                  & ~~50 \\
                 & Nobeyama                 & ~~48 \\
                 & Takahagi                 & ~~50 \\
                 & Yamaguchi                & ~~50 \\
Korea            & KVN                      &  500 \\
                 & Sejong                   &  100 \\
\bottomrule
\end{tabular}
}
\end{table}

Table~\ref{tbl:EAVN Statistics} represents the total observing time of EAVN for each semester.
EAVN offers opportunities of a VLBI observation to astronomers all over the world from September 1 to June 15 with 2.5 months of the system maintenance, while the term of observation shown above is divided into two semesters from September 1 to January 15 next year (the `B' semester) and from January 16 to June 15 (the `A' semester).
As shown in Table~\ref{tbl:EAVN Statistics}, EAVN is providing the observing time of about 1100 hours during two adjacent semesters.
The observing session of EAVN is basically allocated every two weeks with the total observing term of one week each.
EAVN thus provides opportunities of VLBI observations with the total observing duration of two weeks per month from September to June, {while note that EAVN observations are not always allocated to whole observing term tightly since the total observing time provided by each telescope is limited, as shown in Table~\ref{tbl:EAVN Telescope Time}, and target sources tend to concentrate on a particular right ascension range.
These result in the total observing time for a year ($\sim$ 1100~h) shown in Table~\ref{tbl:EAVN Statistics} being much smaller than the total potential observing time ($\sim$ 3000~h) of EAVN.}

%%%%%%%%%%%%%%%
%%% Table 6 %%%
%%%%%%%%%%%%%%%

\begin{table}[H]
\caption{Total observing time of EAVN for each semester.
\label{tbl:EAVN Statistics}}
\tablesize{\endnotesize}

\begin{adjustwidth}{-\extralength}{0cm}
%\centering %% If there is a figure in wide page, please release command \centering
\setlength{\tabcolsep}{2.6mm}{\begin{tabular}{ccccccccc}
\toprule
\textbf{Semester} & \textbf{GOT \boldmath{$^a$} [h]} & \textbf{Session} & \textbf{ToO \boldmath{$^b$} [h]} & \textbf{Session} & \textbf{PE \boldmath{$^c$} [h]} & \textbf{Session} & \textbf{Total Time [h]} & \textbf{Total Session}  \\
\midrule
2018B             & ~~487.5              & ~~78             & ---                  & ---              & ~~32.0            & ~~2              & ~~519.5                 & ~~80 \\
2019A             & ~~439.5              & ~~66             & ~~116.0              & 16               & ~~75.0            & ~~7              & ~~630.5                 & ~~89 \\
2019B             & ~~472.0              & ~~70             & ~~~~~~0.0            & ~~0              & ~~55.0            & ~~6              & ~~527.0                 & ~~76 \\
2020A             & ~~539.5              & ~~79             & ~~~~24.0             & ~~3              & ~~63.0            & ~~6              & ~~626.5                 & ~~88 \\
2020B             & ~~327.5              & ~~58             & ~~~~44.0             & 10               & ~~71.5            & 10               & ~~443.0                 & ~~78 \\
2021A             & ~~581.0              & ~~94             & ~~~~16.0             & ~~2              & ~~26.5            & ~~5              & ~~623.5                 & 101  \\
2021B             & ~~483.5              & ~~72             & ~~~~15.5             & 10               & ~~49.0            & ~~7              & ~~548.0                 & ~~89 \\
2022A             & ~~522.0              & ~~84             & ~~~~13.0             & 11               & ~~34.0            & ~~2              & ~~569.0                 & ~~97 \\
\midrule
Total             & 3852.5               & 601              & 228.5                & 52               & 406.0             & 45               & 4481.0                  & 698  \\
\bottomrule
\end{tabular}}
\end{adjustwidth}
\noindent{\footnotesize Note. $^a$ General observing time. $^b$ Target of opportunity. $^c$ Performance evaluation.}
\end{table}

As of the 2022B semester, EAVN open-use program provides opportunities of observations at 6.7, 22, and 43~GHz with single polarization and with the data recording rate of 1~Gbps.
Dual-polarization capability is opened for the open-use program from the 2023A semester.
EAVN also provides opportunities of the dual-frequency simultaneous data recording mode at 22 and 43~GHz for 9 telescopes (KaVA, Nobeyama, and Sejong), as mentioned in Section~\ref{subsec:Frontend}.
The `multi-frequency data recording mode', on the other hand, is also offered as one of the most distinct modes of EAVN in which KVN records the data of not only 22 and 43~GHz but also 86 and 129~GHz simultaneously.
This mode thus enables us to compare the data at different frequencies directly.

From the 2019A semester, EAVN offers a director's discretionary time for target of opportunity (ToO) observations to track transient or unexpected phenomena swiftly.
As of the 2022A semester, we have conducted 52 ToO sessions with the total observing time of 228.5 hours with maser bursts associated with star-forming regions, Galactic black hole binary, and gamma-ray-emitted extragalactic objects being major targets.

Part of EAVN observations were conducted with non-EAVN telescopes in a framework of EAVN open-use program.
As of the 2022A semester, following non-EAVN telescopes have participated in EAVN observation one or more sessions; Hobart 26~m, Tidbinbilla 70~m (Australia), Effelsberg 100~m (Germany), Noto 32~m, Medicina 32~m, Sardinia 64~m (Italy), Badary 32~m, Svetloe 32~m, Zelenchukskaya 32~m (Russia), and Yebes 40~m (Spain).
These are also a good demonstration for promoting global VLBI activities, such as the EATING VLBI or the Global VLBI Alliance, both of which are explained in Section~\ref{sec:Future Growth into the Global VLBI}.

%%%%%%%%%%%%%%%%%
%%% Section 4 %%%
%%%%%%%%%%%%%%%%%

\section{Scientific Accomplishments by EAVN}
\label{sec:Scientific Accomplishments by EAVN}

Thanks to its high angular resolution at multiple frequencies and plenty of observing time of more than 1100 hours in a year, EAVN is a suitable array to conduct high-cadence and long-term VLBI monitoring toward particular targets.
To make full use of the capabilities of EAVN for deriving initial scientific results and for considering future potential science cases with EAVN, three science working groups (SWGs), AGNs, evolved stars, and star-forming regions, were organized in 2011 before launching KaVA's operation and the large observing programs led by these SWGs have been conducted using KaVA and EAVN from 2015 to 2019 by providing about 50\% of whole observing time of KaVA/EAVN.
Galactic Astrometry SWG was also organized in 2017 for promoting astrometric VLBI observations to investigate dynamics of our galaxy.
In this section, let us introduce overview of activities and scientific achievements to date led by these SWGs.

%%%%%%%%%%%%%%%%%%%
%%% Section 4-1 %%%
%%%%%%%%%%%%%%%%%%%

\subsection{Active Galactic Nuclei}
\label{subsec:Active Galactic Nuclei}

Key issues in AGN science are to investigate the accretion process of the matter onto supermassive black holes (SMBHs), and physical processes on the mechanism of jet formation and its collimation and acceleration in the vicinity of SMBHs.
To approach these important subjects, EAVN AGN SWG is focusing on two targets, our Galactic center Sagittarius A$^*$ which contains a SMBH with its BH mass of $\approx 4.0 \times 10^6 M_{\odot}$ (e.g., \cite{Do19,EHT22a}), and a nearby radio galaxy M87 with its BH mass of $\approx 6.5 \times 10^9 M_{\odot}$ (e.g., \cite{Gebhardt11,EHT19b}), both of which are main targets of the EHT project.

\begin{figure}[H]	

\includegraphics[width=6 cm]{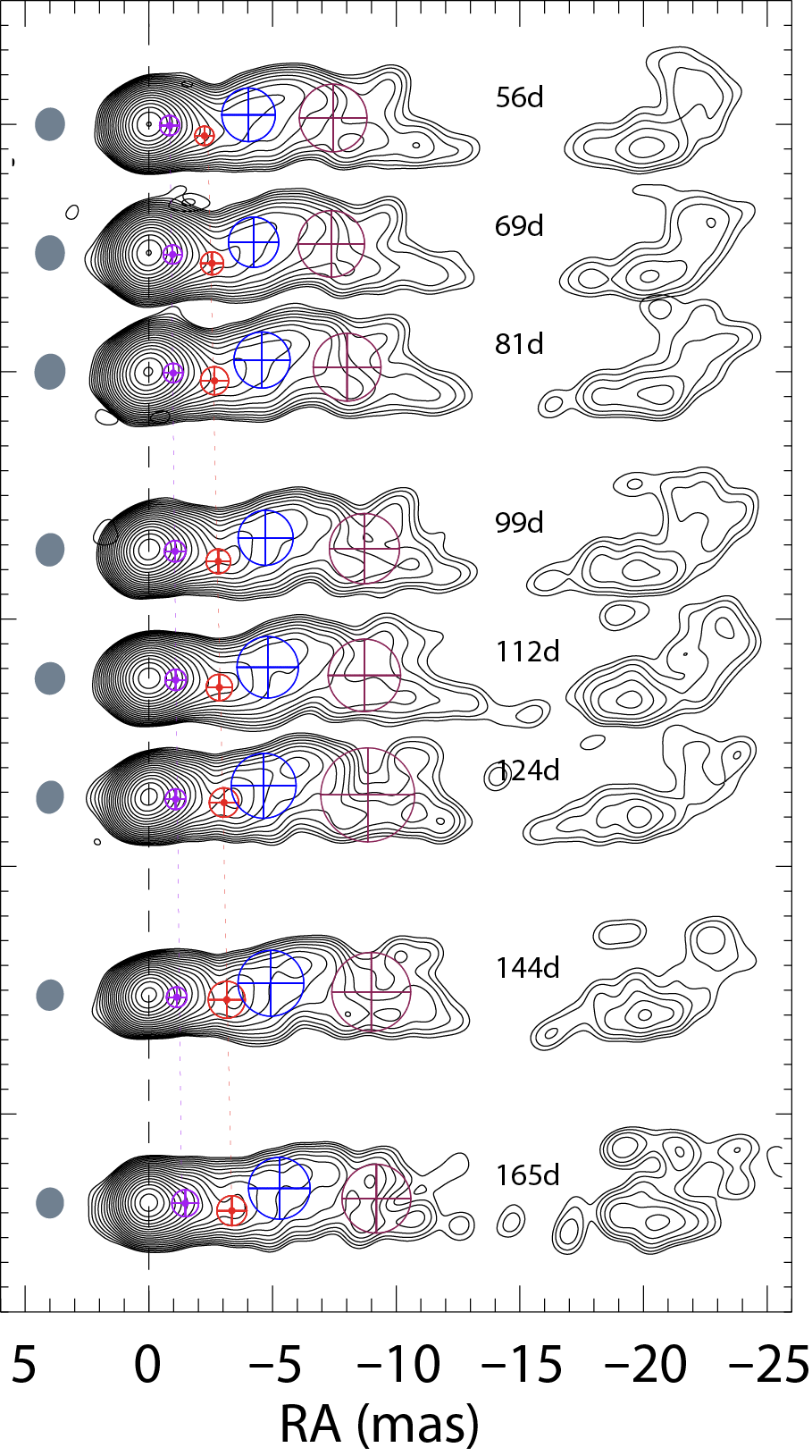}
\includegraphics[width=6 cm]{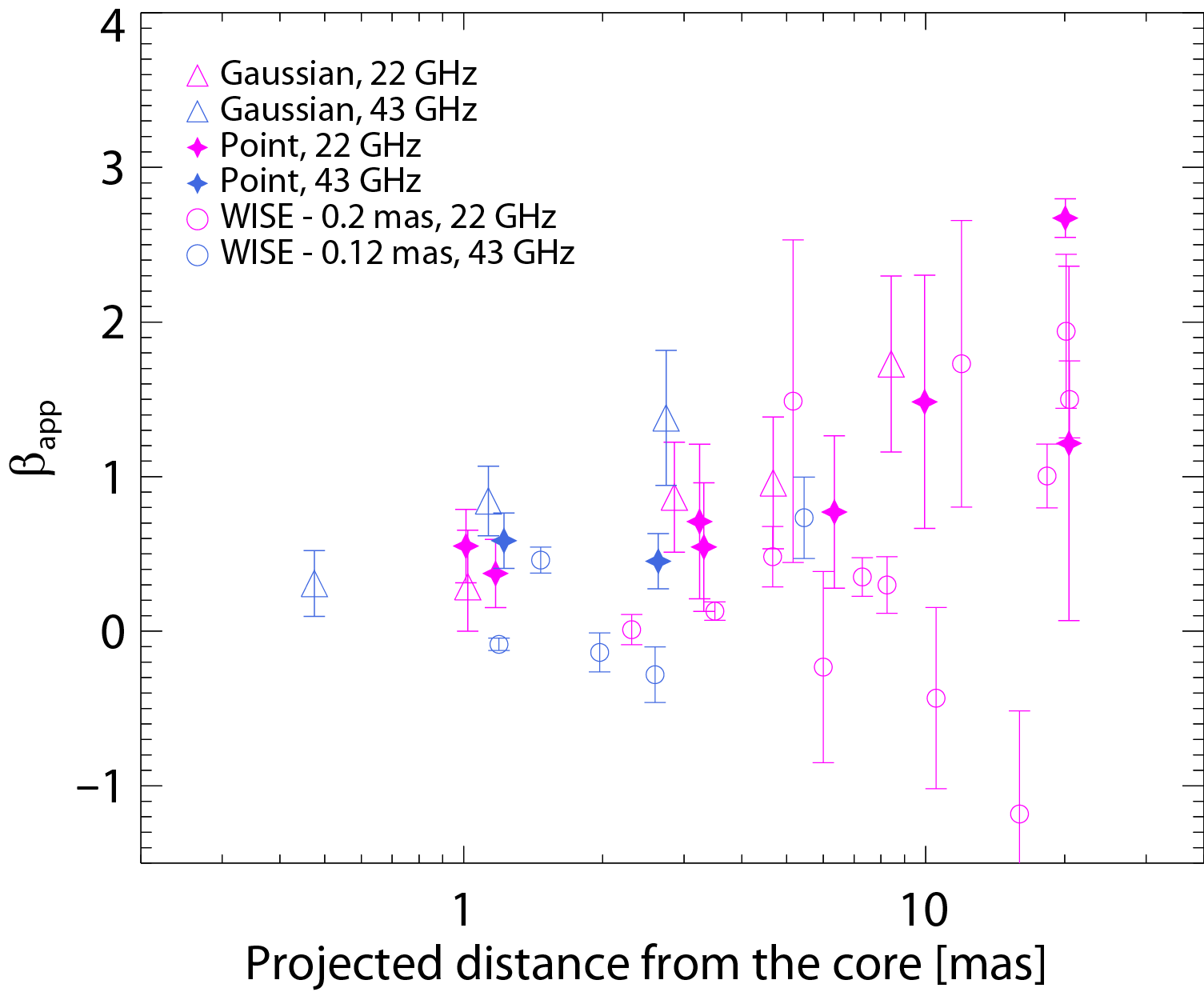}
\caption{({\bf Left}) CLEAN images of M87 obtained with KaVA at 22~GHz.
The images are obtained at eight epochs from 2016 February 26 to June 15.
Circular Gaussian components obtained by the modelfit analysis are overlaid on the images.
({\bf Right}) Apparent jet speed as a function of projected distance from the core of M87.
Symbols shown with magenta and blue colors represent the data obtained at 22 and 43~GHz, respectively.
Adapted from the work in \cite{Park19} with permission from the American Astronomical Society (AAS).
\label{fig:Park_2019_ApJ}}
\end{figure}

The first-epoch observation of M87 was conducted in 2013 using KaVA as a pilot experiment of the large program of KaVA and the monitoring program was continued with the observing interval of two or three weeks.
The first 10-epoch observations in 2013 and 2014 revealed superluminal motions and a trend of gradual acceleration of multiple jet components with its physical scale of 0.1--2~pc, corresponding to 140--2800 Schwarzschild radii \cite{Hada17}.
M87 has been continuously monitored through a large program of KaVA and EAVN since then and those observations revealed the kinematics of the M87 jet on the basis of a detailed velocity profile of the apparent jet speed ($v_{\rm app}$) from the vicinity of the jet base ($\approx 0.5$~mas) with $v_{\rm app} \approx 0.3c$ to $v_{\rm app} \approx 2.7c$ at $\approx 20$~mas from the jet base (\cite{Park19}, see also Figure~\ref{fig:Park_2019_ApJ}).
EAVN observations also reconstructed the fine structure of both approaching and receding jets of M87 with the image dynamic range of around 4000 at both 22 and \mbox{43~GHz \cite{Cui21}}.
EAVN observing campaign of M87 in 2017 spring was carried out as part of quasi-simultaneous multi-wavelength campaign from radio to TeV $\gamma$-ray, which was conducted concurrently with the EHT campaign.
EAVN provided clear images showing knotty jet structure \cite{EHT21}, as well as the flux information at \mbox{cm-wavelengths \cite{EHT19b}.}
Dense monitoring of M87 with five consecutive month in 2016 by KaVA at 22 and 43~GHz quasi-simultaneously could also estimate the magnetic field strength along the jet precisely at an angular scale of 2--10~mas (or 900--4500 Schwarzschild radii for M87) from the core on the basis of detailed determination of the spectral index between two observing \mbox{frequencies \cite{Ro22}}.

%%%%%%%%%%%%%%%%
%%% Figure 4 %%%
%%%%%%%%%%%%%%%%

Another major target of the EHT project, Sgr A$^*$, is also one of the best laboratories for investigating mass accretion processes and time variability of physical properties in the vicinity of SMBH thanks to its proximity.
EAVN AGN SWG is conducting a long-term monitoring of Sgr A$^*$, including an intensive observation program during the EHT campaign every year as another target of the large program of KaVA and EAVN.
Radio emission from Sgr A$^*$ is affected by interstellar scattering effect, resulting in scatter-broadening of its size as the square of the observing wavelength.
This becomes a disadvantage of an observation at a frequency of, e.g., 22~GHz or lower.
Detection of Sgr A$^*$, on the other hand, is also difficult at 43~GHz or higher due to its lower flux density, although Sgr A$^*$ still has a rising spectrum at 43~GHz.
In this point, KaVA is a suitable array for observing Sgr A$^*$ since KaVA can provide a denser ($u$, $v$) coverage corresponding to the significant radio structure thanks to its compact array configuration with the baseline length of 300--2300~km.
Moreover, KaVA has a better angular resolution in the north--south direction at 43~GHz than that of VLBA, thus KaVA can estimate the minor axis size of Sgr A$^*$ more precisely.
On the basis of these advantages, results of precise size measurements of Sgr A$^*$ with its major-axis angular size of $704 \pm 102$~$\upmu$as and $300 \pm 25$~$\upmu$as at 22 and 43~GHz, respectively, can be obtained by KaVA and EAVN \cite{Johnson18,Cho22}.
Those results enable us to compare with the accretion flow dominated model in terms of the distribution of thermal/non-thermal electron and the viewing angle of Sgr A$^*$, both of which can be determined by a theoretical prediction.
EAVN also provided results of the flux density measurement of Sgr A$^*$ at 22 and 43~GHz through the multi-wavelength observation campaign from radio to X-ray, which was conducted concurrently with the EHT campaign in 2017 \cite{EHT22b}.

KaVA/EAVN is participating in an international multi-frequency observing campaign of AGNs besides the EHT campaign mentioned above.
Multi-frequency monitoring of a gravitationally lensed blazar B0218+357, for example, was performed during 2016 and 2020 from radio to GeV $\gamma$-ray and KaVA provided clear images and flux density information at 22 and 43~GHz for the monitoring, in which two lensed images show clear core-jet structure although a distinct jet motion cannot be seen \cite{Hada20,Acciari22}.

KaVA/EAVN is carrying out individual proposal-based observations toward various AGNs as well.
A series of KaVA observations of a nearby bright radio galaxy 3C~84 have revealed the existence and its structure of a circumnuclear plasma disk \cite{Wajima20} and strong jet--cloud interaction on pc-scales \cite{Kino18,Kino21}.
Short- and long-term monitoring observations with KaVA/EAVN show new findings in physical properties of the jet structure and its motion toward several AGNs, e.g., confirming a clear connection between $\gamma$-ray flare and the ejection of a new jet component in the flat spectrum radio quasar 4C+21.35 at 22 and 43~GHz \cite{Lee19}, determining the rebrightened regions in the jet of a nearby radio galaxy NGC~315 at 22~GHz, which is similar to M87 \cite{Park21}, detecting a misaligned jet component in a transient GeV $\gamma$-ray blazar J1544$-$0649 with its maximum viewing angle of 7.4$^{\circ}$ at 6.7~GHz suggesting the misaligned blazar scenario for this source \cite{Shao22}, and finding a tight linear relation between accretion luminosity and jet luminosity in a sample of the {\it Swift} BAT (Burst Alert Telescope)-detected local AGNs with their redshift of smaller than \mbox{0.05 \cite{Baek19}}.
It should also be notable that EAVN is conducting observations toward extragalactic radio sources which are non-traditional VLBI targets, such as a cluster galaxy \cite{Baek22} and a gamma-ray burst object \cite{An20}, thanks to an excellent imaging sensitivity of EAVN although the gamma-ray burst object 190114C could not be detected with EAVN, maybe due to rapid decay of the radio flux.

%%%%%%%%%%%%%%%%%%%
%%% Section 4-2 %%%
%%%%%%%%%%%%%%%%%%%

\subsection{Evolved Stars}
\label{subsec:Evolved Stars}

Evolved stars tend to show strong maser emission of the hydroxyl radical (OH), water vapor (H$_2$O), and silicon monoxide (SiO) at the circumstellar envelope in the asymptotic giant branch (AGB) and post-AGB phases.
Those masers are associated with gas and dust clumps around an evolved star and, thus, a useful tool to reveal the mass loss processes of evolved stars, although its motion is very complex.
To conduct high-cadence VLBI monitoring of a sample of evolved stars for approaching those issues, EAVN Evolved Stars SWG has launched the `ESTEMA' (Expanded Study on Stellar Masers) project using KaVA \cite{Oyadomari16} and the `(renewed) ESTEMA' (EAVN Synthesis of Stellar Maser Animations) project using EAVN.
The most distinguishing strategy of ESTEMA for approaching various scientific objects is to conduct (quasi-)simultaneous observation at 22, 43, 86, and 129~GHz, corresponding to maser emission H$_2$O maser at 22~GHz, and SiO $J$ = 1--0, 2--1, and 3--2 maser at 43, 86, and 129~GHz, respectively, thanks to a unique capability of multi-frequency simultaneous data reception at KaVA, Sejong, and Nobeyama.

\begin{figure}[H]	

\includegraphics[width=6.7 cm]{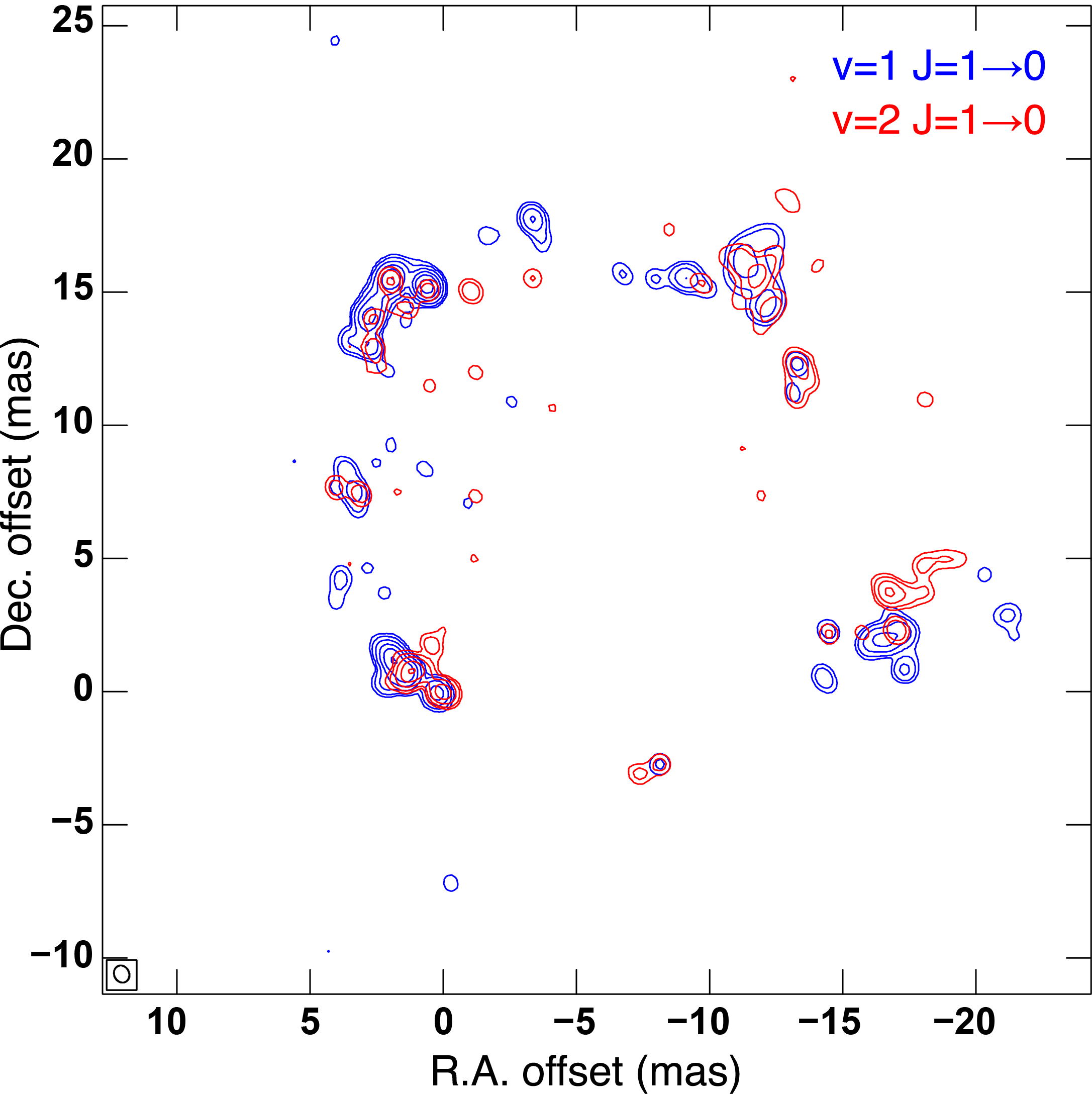}
\includegraphics[width=6.8 cm]{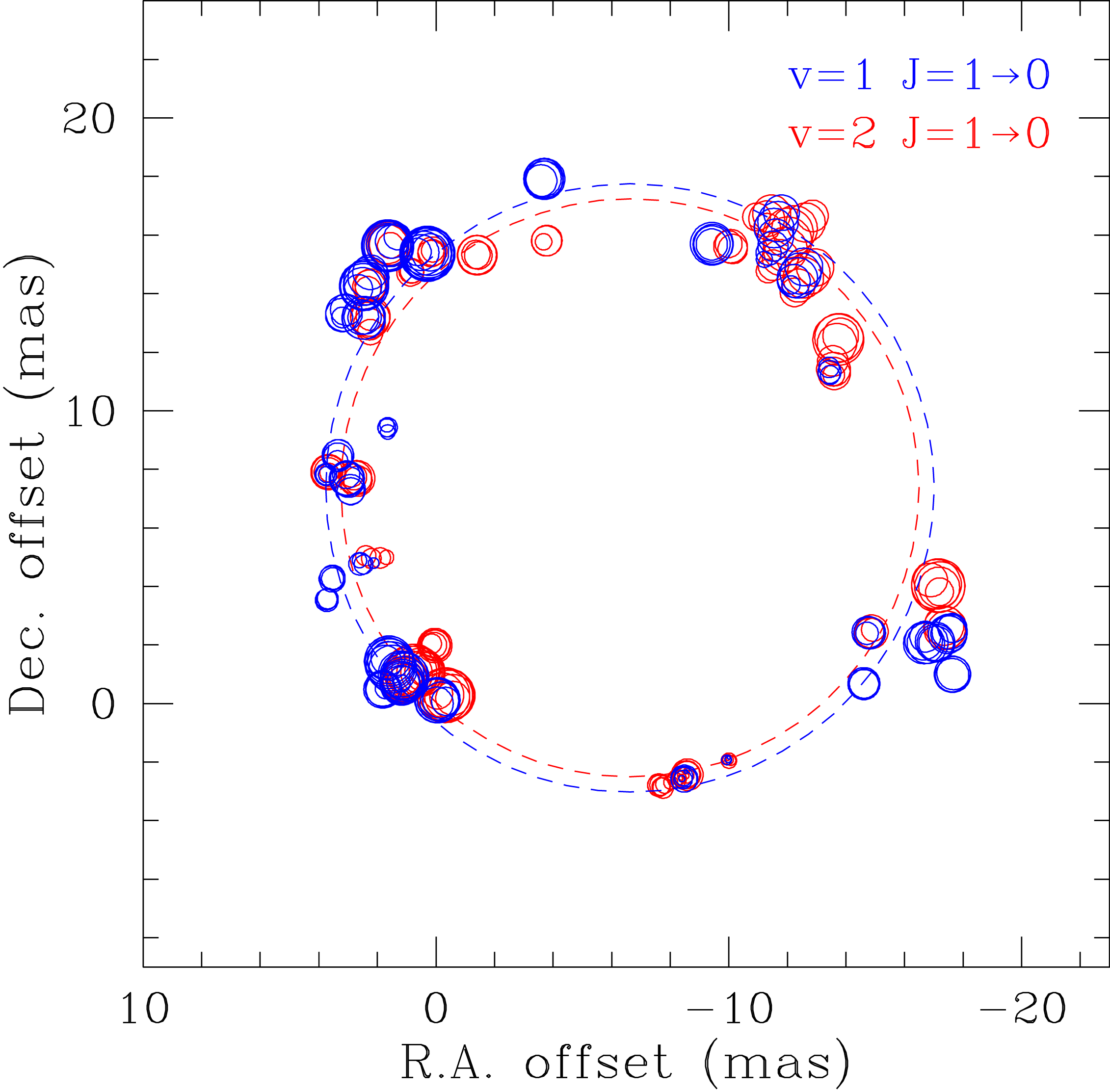}
\caption{({\bf Left}) Spatial distribution of the SiO masers around WX Psc on 2012 April 1.
Blue and red colors represent the $v = 1$ and 2, $J$ = 1--0 SiO maser.
({\bf Right}) Superposed map of SiO masers for $v = 1$ and 2, $J$ = 1--0 in blue and red colors, respectively.
Dashed circles show the average size of the ring-like structure for each SiO maser spot.
Adapted from the work in \cite{Yun16} with permission from AAS.
\label{fig:Yun_2016_ApJ}}
\end{figure}

ESTEMA aims to produce animations of circumstellar H$_2$O and SiO masers toward 10 variable stars by dense monitoring during its pulsation period ($P$) of 1--3 years to investigate three-dimensional kinematics of maser gas clumps, while they are focusing on two objects, a Mira variable BX Cam ($P = 486$~days) and a red supergiant NML Cyg ($P \sim 1000$~days), as a first step.
Prior to launch of ESTEMA, pilot observations of several evolved stars showing distinct maser features at different maser species were conducted using KaVA.
For one of those samples, an OH/IR star WX Psc, single-epoch KaVA observation could reconstruct superposed images of the $v = 1$ and 2, $J$ = 1--0 SiO masers with keeping its relative position by employing astrometric analysis.
Those images clearly show a typical ring-like structure in both SiO masers and the difference in their distribution with $v = 2$ masers being located in an inner region of about 0.5~mas compared to $v = 1$ masers (\cite{Yun16}, see also Figure~\ref{fig:Yun_2016_ApJ}).
As for one of the major targets of ESTEMA, BX Cam, long-term monitoring with EAVN from 2018 May to 2021 June, including 37 epochs clearly described three-dimensional kinematics of H$_2$O masers associated with the circumstellar envelope with providing the animation of H$_2$O masers.
These results help us to understand the mechanism of the propagation of shock waves around evolved stars \cite{Xu22}.

Long-term monitoring of those masers will be a good tool for revealing the physical environments of the circumstellar environment and the pumping mechanism of SiO maser.
Simultaneous observations of H$_2$O and SiO masers at four frequencies (22, 43, 86, and 129~GHz) toward several evolved stars, which are not included in the target of ESTEMA, have also conducted using KVN, with EAVN Evolved Stars SWG being collaborated \mbox{(e.g., \cite{Yoon18,Kim19,Yang20}).}

%%%%%%%%%%%%%%%%
%%% Figure 5 %%%
%%%%%%%%%%%%%%%%

Some of evolved stars in AGB or post-AGB phases are identified as showing fast bipolar outflows traced by H$_2$O masers, called water fountain sources (e.g., \cite{Imai02}).
EAVN Evolved Stars SWG is carrying out follow-up KaVA observations of a sample of ESTEMA targets in which high-velocity components of H$_2$O masers are detected by Nobeyama 45~m telescope in order to confirm the spatial distribution of H$_2$O masers.
They found that H$_2$O masers in IRAS~18286$-$0959, one of the ESTEMA targets, show a symmetric spatial distribution with extremely high expansion velocities of greater than 200~km~s$^{-1}$, suggesting that H$_2$O masers are associated with the bipolar outflow \cite{Imai20}.

%%%%%%%%%%%%%%%%%%%
%%% Section 4-3 %%%
%%%%%%%%%%%%%%%%%%%

\subsection{Star-Forming Regions}
\label{subsec:Star-Forming Regions}

To reveal the process of massive star formation with its mass of greater than 8$M_{\odot}$ is an important issue for understanding basic astrophysics, such as formation of heavy elements and galaxy formation, while it is still poorly understood mainly because of large distance from us and relatively short evolutionary timescale.
Large number of high-mass young stellar objects (HMYSOs) are known to associate strong emission of class I and II methanol (CH$_3$OH) masers at 44 and 6.7~GHz, respectively, and water vapor (H$_2$O) maser at 22~GHz.
Those masers are associated with various scales of structures in HMYSOs, such as outflows, disks, and H\,II regions, VLBI is thus one of the best tools for investigating the process related to high-mass star formation since VLBI can directly obtain images of HMYSOs and trace three-dimensional motions of masers.

EAVN Star-Forming Regions SWG has started large program for long-term monitoring of H$_2$O masers at 22~GHz and class I CH$_3$OH masers at 44~GHz for a sample of HMYSOs containing 25 primary candidates \cite{Kim18}.
The project was conducted mainly using KaVA since target masers tend to resolve with longer baseline lengths by EAVN than that by KaVA, instead the project contains follow-up observations with JVN, VERA, JVLA (Karl G.\ Jansky Very Large Array), and ALMA (Atacama Large Millimeter/submillimeter Array).
Prior to launch of EAVN, imaging test observations of EAVN (VERA + JVN + Sheshan) at 6.7~GHz had been conducted toward 36 HMYSOs which emit class II CH$_3$OH masers in 2010 and 2011.
These observations provide us a basic information on maser distribution and its morphology (elliptical, arched, linear, paired, and complex) for a sample of \mbox{HMYSOs \cite{Fujisawa14,Sugiyama16}.}
It is also notable that a KaVA observation of a high-mass star-forming region IRAS~18151$-$1208 detected class I CH$_3$OH methanol masers with the baseline lengths of shorter than 650~km, suggesting that the maser component has an extended structure with its angular size of larger than 4~mas or $\sim$12~AU in the linear size of the source \cite{Matsumoto14}.
This is the first detection of class I CH$_3$OH maser at 44~GHz with VLBI, indicating that KaVA is a powerful instrument to study physical properties of HMYSOs.

\begin{figure}[H]	
\includegraphics[width=10 cm]{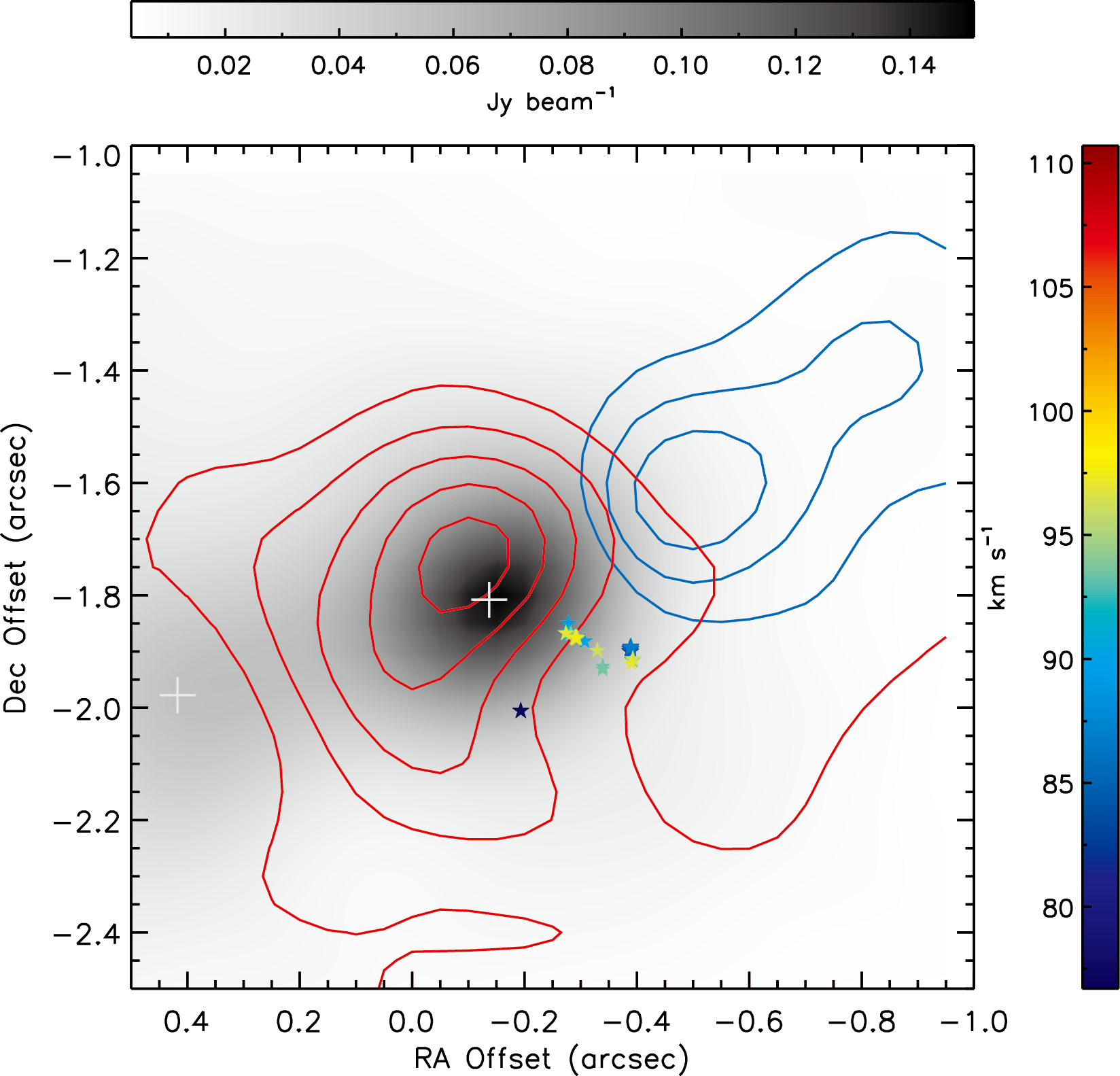}
\caption{Spatial distribution of H$_2$O masers obtained with KaVA, which are shown in colored points.
Blue and red contours indicate the integrated intensity map of the SiO 5--4 line and and gray scale represents the continuum emission at 230~GHz, both of which were obtained with ALMA Band 6.
Adapted from the work in \cite{Kim20} with permission from AAS.
\label{fig:Kim_2020_ApJ}}
\end{figure}

Another good example for studying the process of high-mass star formation with KaVA and ALMA was demonstrated toward a high-mass cluster-forming region G25.82$-$0.17, detecting continuum emission showing a velocity gradient indicating a rotating disk by ALMA at 230~GHz and strong H$_2$O maser sources associated with the root of bipolar outflows in the continuum source by KaVA at 22~GHz.
These results suggest a similar star-formation process between low-mass and high-mass stars (\cite{Kim20}, see also Figure~\ref{fig:Kim_2020_ApJ}).
KaVA also provides precise measurements of proper motion of H$_2$O masers associated with bipolar outflows toward several HMYSOs and high-mass star-forming regions \mbox{(e.g., \cite{Chibueze21,Trinidad21})} thanks to a capability of dense monitoring by KaVA.
These results are of help to us for understanding basic physical processes of HMYSOs.

%%%%%%%%%%%%%%%%
%%% Figure 6 %%%
%%%%%%%%%%%%%%%%

%%%%%%%%%%%%%%%%%%%
%%% Section 4-4 %%%
%%%%%%%%%%%%%%%%%%%

\subsection{Astrometry}
\label{subsec:Astrometry}

Precise astrometric measurement for detecting parallax is one of the most distinct features in VLBI thanks to its high angular resolution \cite{Reid14}.
In East Asia, various studies on measurement of distance to Galactic H$_2$O and SiO maser sources, and the three-dimensional structure and dynamics of the Milky Way have been accomplished (\cite{VERA20} and references therein) based on astrometric measurements with VERA, which is a dedicated VLBI array for Galactic astrometry with the unique capability of the dual-beam system \cite{Honma03}.
On the basis of accomplishments made mainly by VERA, EAVN Galactic Astrometry SWG examined the capability of astrometric measurements and EAVN opened the astrometry mode for measuring parallax using KaVA 7 telescopes for the open-use program in the 2019B semester.
Station coordinates of KaVA have been monitored by 22-GHz geodetic VLBI observations since 2014 and individual baseline lengths are determined with an accuracy of $\sim$3~mm \cite{Jike18,Xu21}.
For atmospheric calibration of KaVA astrometry mode (i.e., estimation of zenith wet excess path delays), the data provided by GPS and JMA (Japan Meteorological Agency meso-scale analysis data for numerical weather prediction\endnote{\url{https://www.jma.go.jp/jma/jma-eng/jma-center/nwp/outline2019-nwp/index.htm}, accessed on 1 December 2022}) are applied.
Those examinations result in suppressing the tropospheric zenith delay residual ($c\Delta \tau_{\rm{trop}}$) of less than $\sim$2~cm \cite{Nagayama15}.

\begin{figure}[H]
\includegraphics[width=10 cm]{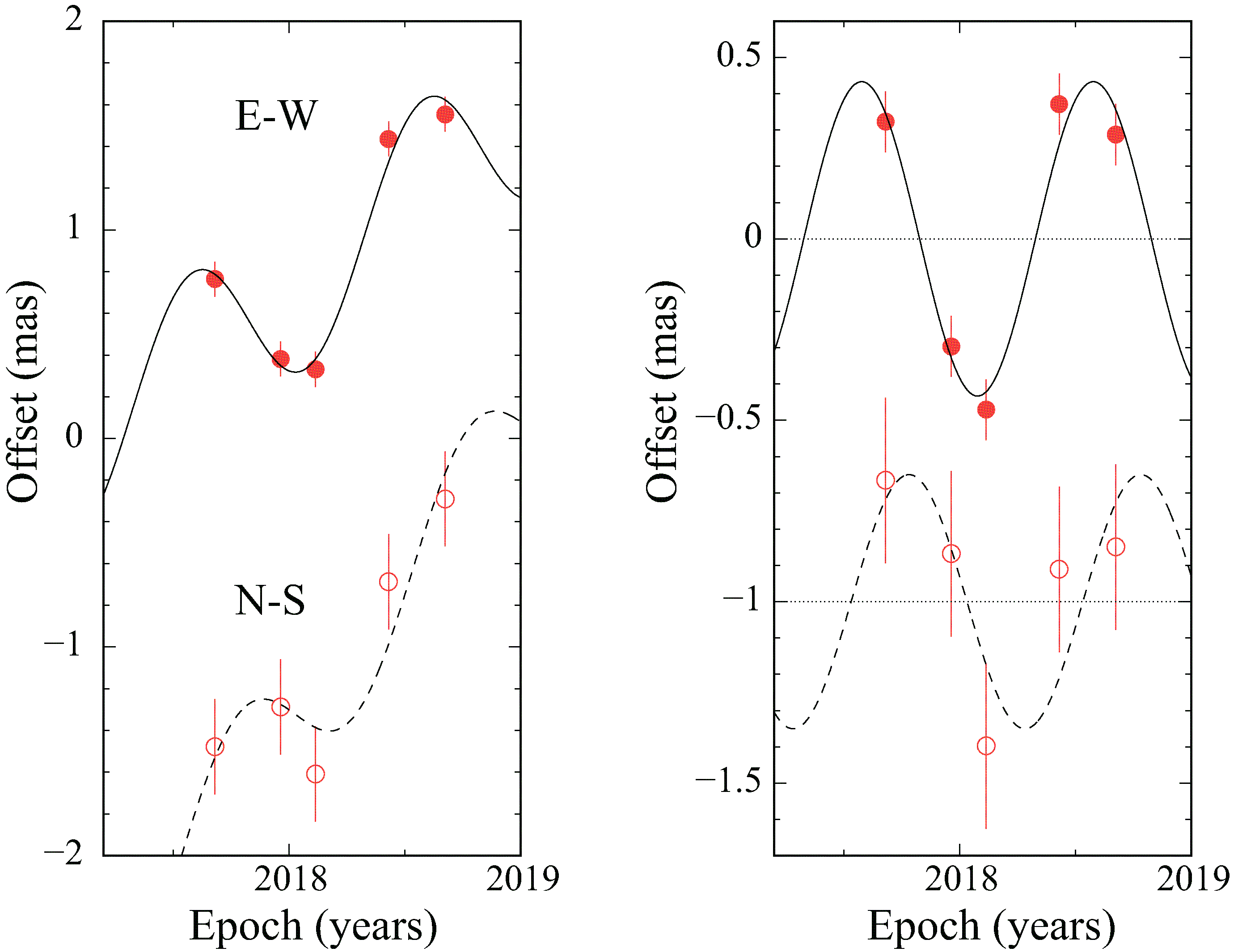}
\caption{Results of parallax measurement and proper-motion fitting for a star-forming region W3(OH). The position offsets of W3(OH) are plotted with respect to the background quasar J0244+6228 in the east ($\alpha$cos$\delta$) and north ($\delta$) directions as a function of time. For clarity, the north direction data is plotted offset from the east direction data. 
({\bf Left}) The best-fit models in the east and north directions are shown as continuous and dashed curves, respectively. ({\bf Right}) Same as the left panel, but the component of proper motion is removed.
\label{fig:EAVN_Parallax}}
\end{figure}

To demonstrate the capability of the astrometry mode, Galactic trigonometric parallax measurement of 22~GHz H$_2$O maser sources associated with a famous star-forming region W3(OH) was conducted with KaVA.
The resultant parallax is measured to be \mbox{$0.460 \pm 0.035$~mas}, which is consistent with a result of $0.489 \pm 0.017$~mas measured by VLBA \cite{Hachisuka06} within the margin of error (see Figure~\ref{fig:EAVN_Parallax}).
More than 10 proposals with the total observing time of greater than 300 hours have already been accepted for astrometric study with KaVA, which covers a wide range of scientific topics including studies of the structure of extreme outer Galaxy \cite{Sakai22}, microquasars, jet swinging in AGN, astrometry of radio stars, and geodetic VLBI observations with simultaneous data reception at 22 and 43~GHz. 

%%%%%%%%%%%%%%%%
%%% Figure 7 %%%
%%%%%%%%%%%%%%%%

%%%%%%%%%%%%%%%%%
%%% Section 5 %%%
%%%%%%%%%%%%%%%%%

\section{Future Growth into the Global VLBI}
\label{sec:Future Growth into the Global VLBI}

As mentioned in Section~\ref{sec:Introduction}, higher angular resolution can be obtained by employing longer baseline length between each telescope.
It is also important to expand the array capability of EAVN by, for example, inviting telescopes and existing VLBI arrays located near and far from EAVN.
International collaboration in VLBI is thus a natural extension to obtain better array capability and we enhance international collaboration in VLBI research with non-EAVN facilities as shown below as examples.

%%%%%%%%%%%%%%%%%%%
%%% Section 5-1 %%%
%%%%%%%%%%%%%%%%%%%

\subsection{Collaboration with Southeast Asian Countries}
\label{subsec:Collaboration with Southeast Asian Countries}

Radio telescope facilities and the radio astronomy community are spreading not only in East Asia but also in Southeast Asia.
In Thailand, construction of 40 m radio telescope (Thai National Radio Telescope; TNRT) has been completed in Chiang Mai led by National Astronomical Research Institute of Thailand (NARIT) and will be operational at 1.0--1.8 and 18.0--26.5~GHz in 2022, while an observable frequency range will be expanded to be 300~MHz--115~GHz in the future \cite{Jaroenjittichai18}.
Key science topics to be achieved with TNRT are summarized in the white paper \cite{Jaroenjittichai22}.
Two new radio telescopes with {their} diameter of 13~m are also under construction in Chiang Mai and Songkhla, both of which conform to the specifications of the VLBI Global Observing System (VGOS).
For another 32 m telecommunication antenna at the campus of CAT Telecom Public Company Limited in Chonburi, the conversion into a radio telescope is also being considered.
These telescopes will constitute the Thai VLBI Network (TVN) with the longest baseline length of 1330~km.
Indonesia plans to convert a 32 m telecommunication antenna into radio telescope.
A new 7 m radio telescope is operated by the Universiti Pendidikan Sultan Idris (UPSI)--Universiti Malaya (UM) Radio Astronomical Observatory in Malaysia.
These are potential telescopes of EAVN for developing array capabilities thanks to its unique location.

For future development in radio astronomy and VLBI in Southeast Asian countries and for close collaboration in these fields between EAVN and those countries, a new memorandum of understanding (MoU) for EAVN has been concluded on 2021 October 1 by inviting NARIT, as well as Yunnan Observatories (YNAO) in China, and National Geographic Information Institute (NGII) in Korea, as a new member institute of EAVN.
This MoU is a renewal of existing one which had been concluded on 2018 September 6 between KASI, NAOJ, Shanghai Astronomical Observatory (SHAO), and Xinjiang Astronomical Observatory (XAO), and will accelerate further practical collaboration in wide fields of VLBI sciences in Asian region.

%%%%%%%%%%%%%%%%%%%
%%% Section 5-2 %%%
%%%%%%%%%%%%%%%%%%%

\subsection{EATING VLBI}
To further connect EAVN to global baselines, let us introduce an example of a practical collaborative VLBI project entitled `East Asia To Italy: Nearly Global VLBI' (EATING VLBI).
Project overview and initial scientific results of EATING VLBI is provided in another review paper in this Special Issue.

EATING VLBI was launched as a trilateral collaborative project between Italy, Japan, and Korea in 2012 and became a regular session of KVN open-use program in 2020.
Current EATING VLBI consists of three Italian telescopes, Medicina 32~m, Noto 32~m, and Sardinia 64~m, and EAVN telescopes shown in Table~\ref{tbl:EAVN-List}, while it operates with the observing frequency of 22~GHz which is a common one for all Italian telescopes and many EAVN telescopes.
The maximum baseline length of EATING VLBI is composed of Sardinia and VERA-Ogasawara stations with its baseline length of 11164~km, corresponding to an angular resolution of 0.25~mas at 22~GHz.
Although a big geographical gap in a ($u$, $v$) plane appears for the original trilateral EATING VLBI, better image sensitivity and ($u$, $v$) coverage can be obtained for renewed EATING VLBI thanks to a big telescope and a unique location in China, such as Tianma 65~m and Nanshan 26~m, respectively.

As of the first half of 2022, EAVN has conducted EATING VLBI observations with the number of sessions of 37 and with the total observing time of 275 hours since the launch of EAVN open-use program in 2018.
About 49\% (135 hours) of those observations involved any of Italian telescopes so that they can secure the common sky coverage with EAVN.
It should also be noted that part of the EATING VLBI sessions were conducted in conjunction with the Quasar network (Badary 32~m, Svetloe 32~m, and Zelenchukskaya 32~m) in Russia \cite{Shuygina19}.
Major targets of EATING VLBI conducted to date are M87 and the nearest gamma-ray emitting narrow-line Seyfert 1 (NLS1) galaxy 1H~0323+342.
It is notable that a series of EATING VLBI observations could provide images with the highest angular resolution to date toward 1H~0323+342 compared to previous studies with VLBI (e.g., \cite{Wajima14,Fuhrmann16,Doi18,Hada18}), which enable us to investigate a gamma-ray emission mechanism of NLS1.
Imaging capability and initial scientific results of these targets are shown in another review paper of this Special Issue.

Three Italian telescopes are planning installation of a new compact triple-band receiver (CTR) \cite{Han17}, in which simultaneous reception of radio signal at 22, 43, and 86~GHz is available.
Design of the quasi-optics and assembly of CTR is carried out at KASI and installation to each telescope will be conducted in 2023.

%%%%%%%%%%%%%%%%%%%
%%% Section 5-3 %%%
%%%%%%%%%%%%%%%%%%%

\subsection{Global VLBI Alliance}

The open-use program is operated by a lot of domestic and international VLBI arrays besides EAVN, such as EVN, LBA, and VLBA, as mentioned in Section~\ref{sec:Introduction}.
It is thus a natural extension of VLBI array capability by integrating these arrays into a single VLBI array to obtain 100~$\mu$as-scale angular resolution at cm- to mm-wavelengths.
This motivates us to set up an idea of the Global VLBI Alliance (GVA)\endnote{See also the following site for more details of GVA: \url{http://gvlbi.evlbi.org/}, accessed on 1 December 2022} \cite{Colomer21}.
GVA aims not only providing an opportunity of obtaining access to the global VLBI array but also sharing various knowledge related to VLBI, such as scientific strategies, technical developments between each VLBI array, and discussion on observational campaigns using GVA.
The official working group for GVA has already been established under the commission B4 (radio astronomy) of the International Astronomical Union in 2020\endnote{\url{https://www.iau.org/science/scientific_bodies/working_groups/324/}, accessed on 1 December 2022} and started practical discussion with four VLBI arrays shown above (EAVN, EVN, LBA, and VLBA) being currently a regular member \mbox{of GVA}.

%%%%%%%%%%%%%%%%%
%%% Section 6 %%%
%%%%%%%%%%%%%%%%%

\section{Conclusions}
\label{sec:Conclusions}

The paper describes overview of the observing system of the East Asian VLBI Network (EAVN) and brief summary of scientific accomplishments to date with EAVN.
EAVN is currently operated using 16 radio telescopes and three correlators located in China, Japan, and Korea, with the longest baseline length of 5078~km, corresponding to the highest angular resolution of 1.82, 0.55, and 0.28~mas at 6.7, 22, and 43~GHz, respectively.
EAVN is offering a total observing time of about 1100 hours in a year for the open-use program and occasional observations using the director's discretionary time to astronomers all over the world.
Although EAVN is operated with limited number of observing frequencies, EAVN provides opportunities of VLBI observations with dual-frequency data receiving system at 22 and 43~GHz, which is one of the most unique capabilities of EAVN, enabling us to obtain better baseline sensitivity and to determine a relative source position by employing the frequency phase transfer technique and the source-frequency phase-referencing technique, respectively.
EAVN conducts VLBI observations toward various Galactic and extragalactic radio sources, while EAVN provides observing opportunities to three major scientific objectives, active galactic nucleus, evolved star, and star-forming region, through the large program of KaVA/EAVN.
EAVN will become one of the most important VLBI facilities for future global VLBI thanks to its unique location and unique capability of multi-frequency simultaneous data receiving system.

%%%%%%%%%%%%%%%%%%%%%%%%%%%%%%%%%%%%%%%%%%
\vspace{6pt}

%%%%%%%%%%%%%%%%%%%%%%%%%%%%
%%% Author Contributions %%%
%%%%%%%%%%%%%%%%%%%%%%%%%%%%

\authorcontributions{{The corresponding} author (K.W.) wrote all the manuscript.
Each author contributes to EAVN operation and scientific activities as shown below:
EAVN director (M.H., K.-T.K., Z.-Q.S., N.W.), EAVN AGN Science Working Group (K.A. (Kazunori Akiyama), J.-C.A., T.A., K.A. (Keiichi Asada), Z.C., X.C., I.C., Y.C., A.D., J.D., W.G. (Wei Gou), W.G. (Wen Guo), K.H., Y.H., J.A.H., Y.J. (Yongbin Jiang), Y.J. (Yongchen Jiang), N.K., M.K., S.K., J.-W.L., J.A.L., S.-S.L., B.L., X.L. (Xiaofei Li), R.-S.L., M.N., K.N., J.O., T.O., J.P., H.R., S.S.-S., B.W.S., Y.S., M.T., F.T., S.T., J.W., X.W., W.Y., K.Y., L.Y., Y.Z., G.-Y.Z., R.Z., W.Z.), EAVN Evolved Stars Science Working Group (J.O.C., S.-H.C., H.I., D.-J.K., J.K., Y.Y. (Youngjoo Yun)), EAVN Star-Forming Regions Science Working Group (K.A. (Kitiyanee Asanok), D.-Y.B., T.H., K.S.), EAVN Galactic Astrometry Science Working Group (T.J. (Taehyun Jung), S.-W.K., C.O., N.S., S.X., B.Z.), management and operation of the Korea-Japan Correlation Center (D.-K.J., H.-R.K., S.-J.O., D.-G.R., J.-H.Y.), EAVN site operation and management (W.C., H.-S.C., L.C., K.F., W.J., T.J. (Takaaki Jike), J.-S.K., H.K., S.W.L., G.L., Z.L., Q.L., X.L. (Xiang Liu), K.M., H.O., K.M.S., Y.T., S.W., B.X., H.Y. (Hao Yan), S.-O.Y., Y.Y. (Yoshinori Yonekura), H.Y. (Hasu Yoon), J.Y., H.Z.), collaboration between EAVN and Southeast Asia (T.C., P.J., B.H.K., S.P., W.R., B.S.), overall coordination of EAVN (K.W.). All authors have read and agreed to the published version of the manuscript.
}

%%%%%%%%%%%%%%%
%%% Funding %%%
%%%%%%%%%%%%%%%

\funding{
This work is funded by following: JSPS (Japan Society for the Promotion of Science) Grant-in Aid for Scientific Research (KAKENHI) (S) 18H05222 (T.H.), (A) 16H02167 (H.I.), 18H03721 (K.N.), 21H04524 (H.I.), 22H00157 (K.H.), (B) 18KK0090 (K.H.), 21H01120 (Y.Y.\ (Yoshinori Yonekura)), (C) 17K05398 (T.H.), 19K03921 (K.S.), 21K03628 (S.S-S.),
JSPS Grant-in-Aid for Scientific Research on Innovative Areas (Research in a proposed research area) 21H00032 (Y.Y.\ (Yoshinori Yonekura)), 21H00047 (H.I.),
JSPS Grant-in-Aid for Transformative Research Areas (A) 20H05845 (T.H.),
the Grant of PIIF Heiwa Nakajima Foundation in 2019 (K.S.),
and the Mitsubishi Foundation 201911019 (K.H.).
K.A.\ (Kazunori Akiyama) is financially supported by grants from the National Science Foundation (AST-1935980, AST-2034306, AST-2107681, AST-2132700,  OMA-2029670).
L.C.\ is supported by the Chinese Academy of Sciences (CAS) 'Light of West China' Program (No.\ 2021-XBQNXZ-005) and the National Science Foundation of China (NSFC; No.\ U2031212 and 61931002).
W.C.\ is supported by NSFC (No.\ 11903079).
Z.C.\ is supported by NSFC (No.\ U1931135).
W.J.\ is supported by NSFC (No.\ 12173074).
R.-S.L.\ is supported by the Max Planck Partner Group of the MPG and the CAS, the Key Program of the NSFC (No.\ 11933007), the Key Research Program of Frontier Sciences, CAS (No.\ ZDBS-LY-SLH011), and the Shanghai Pilot Program for Basic Research - CAS, Shanghai Branch (No.\ JCYJ-SHFY-2022-013).
S.T.\ acknowledges financial support from the National Research Foundation of Korea (NRF) through grant no.\ 2022R1F1A1075115.
B.Z.\ is supported by the NSFC (No.\ U1831136 and U2031212) and Shanghai Astronomical Observatory (N-2020-06-09-005).
}

\dataavailability{
All observing data obtained by EAVN (including KaVA), except the data in the term of right of occupation by a principal investigator, are archived via the following website. \\
\url{https://radio.kasi.re.kr/kvn_arch/kvn_search.php}
}

%%%%%%%%%%%%%%%%%%%%%%%
%%% Acknowledgments %%%
%%%%%%%%%%%%%%%%%%%%%%%

\acknowledgments{
This work is made use of the East Asian VLBI Network (EAVN), which is operated under cooperative agreement by National Astronomical Observatory of Japan (NAOJ), Korea Astronomy and Space Science Institute (KASI), Shanghai Astronomical Observatory (SHAO), Xinjiang Astronomical Observatory (XAO), Yunnan Observatories (YNAO), National Geographic Information Institute (NGII), and National Astronomical Research Institute of Thailand (Public Organization) (NARIT), with the operational support by Ibaraki University, Yamaguchi University, and Kagoshima University.
Operation of Hitachi 32~m, Takahagi 32~m, and Yamaguchi 32~m telescopes is partially supported by the inter-university collaborative project `Japanese VLBI Network (JVN)' of NAOJ.}

%%%%%%%%%%%%%%%%%%%%%%%%%%%%%
%%% Conflicts of Interest %%%
%%%%%%%%%%%%%%%%%%%%%%%%%%%%%

\conflictsofinterest{The authors declare no conflicts of interest.}

%%%%%%%%%%%%%%%%%%%%%
%%% Abbreviations %%%
%%%%%%%%%%%%%%%%%%%%%

\abbreviations{Abbreviations}{
The following abbreviations are used in this manuscript:\\

\noindent 
\begin{tabular}{@{}ll}
AGB         & Asymptotic Giant Branch \\
AGN         & Active Galactic Nucleus \\
CTR         & Compact Triple-Band Receiver \\
CVN         & Chinese VLBI Network \\
DAS         & Data Acquisition System \\
DBBC        & Digital Base Band Converter \\
DSP         & Digital Signal Processing \\
EACOA       & East Asian Core Observatories Association \\
EAMA        & East Asian Meeting on Astronomy \\
EATING VLBI & East Asia To Italy: Nearly Global VLBI \\
EAVN        & East Asian VLBI Network \\
EHT         & Event Horizon Telescope \\
EOP         & Earth Orientation Paraeter \\
ESTEMA      & Expanded Study on Stellar Masers / EAVN Synthesis of Stellar Maser Animations \\
EVN         & European VLBI Network \\
FPT         & Frequency Phase Transfer \\
GMVA        & Global Millimeter-VLBI Array \\
%GVA         & Global VLBI Alliance \\
%HINOTORI    & Hybrid Installation Project in Nobeyama, Triple-Band Oriented \\
%HMYSO       & High-Mass Young Stellar Object \\
%ICRF        & International Celestial Reference Frame \\
%ITRF        & International Terrestrial Reference Frame \\
%J-Net       & Japanese VLBI Network \\
%JVN         & Japanese VLBI Network \\
%KaVA        & KVN and VERA Array \\
%KJCC        & Korea-Japan Correlation Center \\
%KNIFE       & Kashima-Nobeyama Interferometer \\
%KVN         & Korean VLBI Network \\
%LBA         & Long Baseline Array \\
%LHCP        & Left-Hand Circular Polarization \\
%RHCP        & Right-Hand Circular Polarization \\
%SEFD        & System Equivalent Flux Density \\
%SFPR        & Source-Frequency Phase-Referencing \\
%SMBH        & Supermassive Black Hole \\
%TDRSS       & Tracking and Data Relay Satellite System \\
%TNRT        & Thai National Radio Telescope \\
%ToO         & Target of Opportunity \\
%TRAO        & Taeduk Radio Astronomy Observatory \\
%TVN         & Thai VLBI Network \\
%VERA        & VLBI Exploration of Radio Astrometry \\
%VGOS        & VLBI Global Observing System \\
%VLBA        & Very Long Baseline Array \\
%VLBI        & Very Long Baseline Interferometry \\
%VSOP        & VLBI Space Observatory Programme
\end{tabular}}

{\noindent \small
\begin{tabular}{@{}ll}
%AGB         & Asymptotic Giant Branch \\
%AGN         & Active Galactic Nucleus \\
%CTR         & Compact Triple-Band Receiver \\
%CVN         & Chinese VLBI Network \\
%DAS         & Data Acquisition System \\
%DBBC        & Digital Base Band Converter \\
%DSP         & Digital Signal Processing \\
%EACOA       & East Asian Core Observatories Association \\
%EAMA        & East Asian Meeting on Astronomy \\
%EATING VLBI & East Asia To Italy: Nearly Global VLBI \\
%EAVN        & East Asian VLBI Network \\
%EHT         & Event Horizon Telescope \\
%EOP         & Earth Orientation Paraeter \\
%ESTEMA      & Expanded Study on Stellar Masers / EAVN Synthesis of Stellar Maser Animations \\
%EVN         & European VLBI Network \\
%FPT         & Frequency Phase Transfer \\
%GMVA        & Global Millimeter-VLBI Array \\
GVA         & \hspace{10mm} Global VLBI Alliance \\
HINOTORI    & \hspace{10mm} Hybrid Installation Project in Nobeyama, Triple-Band Oriented \\
HMYSO       & \hspace{10mm} High-Mass Young Stellar Object \\
ICRF        & \hspace{10mm} International Celestial Reference Frame \\
ITRF        & \hspace{10mm} International Terrestrial Reference Frame \\
J-Net       & \hspace{10mm} Japanese VLBI Network \\
JVN         & \hspace{10mm} Japanese VLBI Network \\
KaVA        & \hspace{10mm} KVN and VERA Array \\
KJCC        & \hspace{10mm} Korea-Japan Correlation Center \\
KNIFE       & \hspace{10mm} Kashima--Nobeyama Interferometer \\
KVN         & \hspace{10mm} Korean VLBI Network \\
LBA         & \hspace{10mm} Long Baseline Array \\
LHCP        & \hspace{10mm} Left-Hand Circular Polarization \\
RHCP        & \hspace{10mm} Right-Hand Circular Polarization \\
SEFD        & \hspace{10mm} System Equivalent Flux Density \\
SFPR        & \hspace{10mm} Source-Frequency Phase-Referencing \\
SMBH        & \hspace{10mm} Supermassive Black Hole \\
TDRSS       & \hspace{10mm} Tracking and Data Relay Satellite System \\
TNRT        & \hspace{10mm} Thai National Radio Telescope \\
ToO         & \hspace{10mm} Target of Opportunity \\
TRAO        & \hspace{10mm} Taeduk Radio Astronomy Observatory \\
TVN         & \hspace{10mm} Thai VLBI Network \\
VERA        & \hspace{10mm} VLBI Exploration of Radio Astrometry \\
VGOS        & \hspace{10mm} VLBI Global Observing System \\
VLBA        & \hspace{10mm} Very Long Baseline Array \\
VLBI        & \hspace{10mm} Very Long Baseline Interferometry \\
VSOP        & \hspace{10mm} VLBI Space Observatory Programme
\end{tabular}}

%%%%%%%%%%%%%%%%%%%%%%%%%%%%%%%%%%%%%%%%%%

%%%%%%%%%%%%%%%%%%
%%% References %%%
%%%%%%%%%%%%%%%%%%

\begin{adjustwidth}{-\extralength}{0cm}
\printendnotes[custom] % Un-comment to print a list of endnotes
\reftitle{References}

\end{adjustwidth}

\end{document}